\documentclass{aa}  

\usepackage{graphicx}
\usepackage{CJKutf8}
\usepackage[varg]{txfonts}
\usepackage{amsmath}	
\usepackage{mathrsfs}
\usepackage{bm}
\usepackage{color}
\usepackage{ulem}
\usepackage{xspace}
\usepackage{threeparttable}
\usepackage[colorlinks=true, linkcolor=blue, citecolor=blue, urlcolor=blue]{hyperref}
\definecolor{mygray}{gray}{0.6}
\definecolor{orange}{rgb}{1.0, 0.4, 0.0}
\definecolor{myblue}{rgb}{0.1, 0.5, 0.7}
\definecolor{mygreen}{rgb}{0.2, 0.6, 0.4}
\newcommand{\ccc}[1]{\textcolor{orange}{[\textit{CWO: \small #1}]}}

\newcommand{\corem}[1]{\textcolor{mygray}{\sout{#1}}}
\newcommand{\cim}[1]{\textcolor{orange}{[\textit{$\star$chris$\star$}:\textbf{\small #1}}]}

\newcommand{\md}{\textbf}

\definecolor{mygray}{gray}{0.6}
\newcommand{\ywc}[1]{[\textcolor[RGB]{128,0,128}{\small YW: \textit{#1}}]} 

\newcommand{\smc}[1]{\textcolor{mygreen}{[\textit{SM: \small #1}]}}

\newcommand{\app}[1]{App.~\ref{sec:#1}}
\newcommand{\Se}[1]{Section~\ref{sec:#1}}
\newcommand{\se}[1]{Sect.~\ref{sec:#1}}
\newcommand{\Fg}[1]{Figure~\ref{fig:#1}}
\newcommand{\fg}[1]{Fig.~\ref{fig:#1}}

\newcommand{\tb}[1]{Table~\ref{tab:#1}}

\newcommand{\eq}[1]{equation~(\ref{eq:#1})}

\newcommand{\eqs}[2]{equations~(\ref{eq:#1}) and (\ref{eq:#2})}

\renewcommand{\ywc}[1]{}
\renewcommand{\smc}[1]{}
\renewcommand{\cim}[1]{}

\renewcommand{\corem}[1]{\xspace}
\renewcommand{\ccc}[1]{\xspace}

\renewcommand{\md}{}

\newcommand\gas{\mathrm{g}}

\newcommand\parti{\mathrm{p}}

\begin{document}

\title{Solving for the 2D Water Snowline with Hydrodynamic Simulations}
\subtitle{Emergence of gas outflow, water cycle and temperature plateau}


\author{Yu Wang(\begin{CJK*}{UTF8}{gbsn}王雨\end{CJK*})\inst{1}
	\and
	Chris W. Ormel\inst{1}
	\and
	Shoji Mori(\begin{CJK*}{UTF8}{min}森昇志\end{CJK*})\inst{2,3}
	\and
	Xue-Ning Bai(\begin{CJK*}{UTF8}{gbsn}白雪宁\end{CJK*})\inst{2,1}
}

\institute{Department of Astronomy, Tsinghua University, 30 Shuangqing Rd, Haidian DS, 100084 Beijing, China\\
	\email{wang-y21@mails.tsinghua.edu.cn}
	\and
	Institute for Advanced Study, Tsinghua University, 30 Shuangqing Rd, Haidian DS, 100084 Beijing, China
	\and
	Astronomical Institute, Graduate School of Science, Tohoku University, 6-3 Aoba, Aramaki, Aoba-ku, Sendai, Miyagi, 980-8578, Japan
}

\date{Received September 15, 1996; accepted March 16, 1997}

\abstract
{
In protoplanetary disks, the water snowline marks the location where inward-drifting, ice-rich pebbles sublimate, releasing silicate grains and water vapor. These processes can trigger pile-ups of solids, making the water snowline a promising site for forming planetesimals, e.g., by streaming instabilities.
	However, previous studies exploring the dust pile-up conditions typically employ 1D, vertically-averaged and isothermal assumptions.}
{In this work, we investigate how the 2D flow pattern and a realistic temperature structure affect the accumulation of pebbles at the snowline and how latent heat effects can leave observational imprints.}
{We perform 2D multifluid hydrodynamic simulations in the disk's radial-vertical plane with \texttt{Athena++}, tracking chemically heterogeneous pebbles and the released vapor. With a recently-developed phase change module, the mass transfer and latent heat exchange during ice sublimation are calculated self-consistently. The temperature is calculated by a two-stream radiation transfer method under various opacities and stellar luminosity.}
{We find that vapor injection at the snowline drives a previously unrecognized outflow, leading to a pile-up of ice outside the snowline. Vapor injection also decreases the headwind velocity in the pile-up, promoting planetesimal formation and pebble accretion.
	In actively-heated disks,
	we identify a water-cycle: after ice sublimates in the hotter midplane, vapor recondenses onto pebbles in the upper, cooler layers, which settle back to the midplane. This cycle enhances the trapped ice mass in the pile-up region. Latent heat exchange flattens the temperature gradient across the snowline, broadening the width while reducing the peak solid-to-gas ratio of pile-ups.
}
{Due to the water cycle, active disks are more conducive to planetesimal formation than passive disks. The significant temperature dip (up to 40\,K) caused by latent heat cooling manifests as an intensity dip in the dust continuum, presenting a new channel to identify the water snowline in outbursting systems. }


\maketitle
%
\section{Introduction}
Planets are born in protoplanetary disks, going through a series of growth from dust grains to planetesimals before the formation of planet embryos and the final assembly of planets \citep{DrazkowskaEtal2023, Birnstiel2024}. Tiny $\mu$m-sized grains, inherited from the molecular cloud, can readily grow by coagulation \citep{DominikTielens1997,OrmelEtal2007}.
However, when reaching cm-sized pebbles, growth is believed to stall due to two key factors.
First, as relative velocities increase with dust size, collisions result in fragmentation or bouncing rather than coagulation \citep{BlumWurm2000,BlumWurm2008}. Second, pebbles experience rapid inward radial drift, which quickly drains the solid mass budget \citep{Weidenschilling1977}.
To overcome these growth barriers, a local enhancement in solids is crucial. An elevated solid-to-gas ratio can trigger streaming instabilities (SI), which in turn induce gravitational collapse of pebbles, leading to direct formation of km-sized planetesimals \citep{YoudinGoodman2005,JohansenEtal2007,JohansenEtal2009}. 

Disk snowlines—particularly the water snowline—are thought to be promising sites for planetesimal formation due to their capacity to enhance solid material \citep{DrazkowskaEtal2023}. This enhancements occurs due to two effects. First, as pebbles drift across the water snowline, their ice contents sublimates, leaving behind pure silicates. This process creates a “traffic jam” effect if the decreasing stickiness of dry silicates results in low fragmentation threshold \citep{DominikTielens1997,WadaEtal2013,GundlachBlum2015} or if ice-rich pebbles disaggregate upon sublimation \citep{SaitoSirono2011,AumatellWurm2011}. Both scenarios would reduce the pebble size and slow down their inward radial drift, leading to a pile-up of silicates-dominated pebbles interior to the snowline \citep{SaitoSirono2011,HyodoEtal2019,HyodoEtal2021}. Second, exterior to the snowline the density in ice can be boosted when vapor diffuses back across it and recondenses onto ice-rich pebbles, creating an additional pile-up \citep{CuzziZahnle2004,RosJohansen2013,SchoonenbergOrmel2017,DrazkowskaAlibert2017}. In both the “traffic jam” and “vapor retro-diffusion” mechanisms, numerical studies have shown that the midplane dust-to-gas ratio can readily reach unity (e.g., \citealt{SchoonenbergOrmel2017,HyodoEtal2019}), a level necessary to trigger SI \citep{JohansenYoudin2007,CarreraEtal2015,BaiStone2010,LiYoudin2021,LimEtal2024}.

However, recently, the viability of the traffic jam effect has been called into question. Laboratory experiments find that ice's stickiness drops sharply below 200 K, suggesting that silicates are not more fragile than ice \citep{GundlachEtal2018,MusiolikWurm2019}. Furthermore, multi-band analysis of ALMA observations towards V883 Ori, an outburst system that has extended its snowline to $\approx$40-80 au, reveals continuous dust sizes across the snowline \citep{HougeEtal2024}. This leaves the vapor retro-diffusion mechanism as the key snowline process to elevate the solid-to-gas ratio.

Previous studies investigating factors conducive to solid pile-up typically employ 1D, vertically-averaged and isothermal assumptions.
Under these assumptions, \citet{SchoonenbergOrmel2017} find that a high pebble flux, low Stokes number and large diffusivity are beneficial to form ice pile-up. Additionally, accounting for the back reaction of solids on the gas can double the enhancement of ice. The ``traffic jam'' of silicates can even act in a runaway fashion when considering the back reaction on diffusivity of dust and gas, which decreases with high solid concentration \citep{HyodoEtal2019}.

However, the vertical structure in the disk in reality plays an important role. First, in addition to radial transport, pebbles settle vertically and vapor can diffuse along the $z$-direction. Thus, recondensation-driven pebble growth could strengthen vertical settling, potentially enhancing pile-ups.
Second, a vertical temperature gradient in the disk will determine the morphology of the snowline. The water snowline in the solar nebula might have resided in the actively-heated regions (e.g., \citealt{OkaEtal2011}, but also see \citealt{MoriEtal2021}), where the midplane is hotter than the upper layer due to viscous dissipation. In contrast, passively-heated disks usually hold cooler midplanes (e.g., \citealt{ChiangGoldreich1997}).
These different vertical temperature structures lead to distinct snowline morphology as illustrated in \citet{OkaEtal2011}: active heating sublimates ice at the midplane, creating a narrow ice-condensing region above the midplane.
\citet{RosJohansen2013} have demonstrated that a narrower ice-condensing region speeds up pebble growth by concentrating the material supply for recondensation.
However, their Monte Carlo simulations, which are conducted in a local box with periodic boundaries in the radial direction, cannot directly address the consequences of snowline morphology to the solid pile-up.

Moreover, ice sublimation and vapor condensation involve latent heat energy exchange with the surrounding environment (e.g., \citealt{Owen2020}). Though typically neglected in disk snowline modeling, latent heat absorption has been shown to significantly shift the sublimation front (i.e., the snowline), creating an ``isothermal'' region in accreting planets' envelopes \citep{WangEtal2023}. \citet{DrazkowskaAlibert2017} shows that different disk temperature profiles lead to different strength of pile-up and a stronger pile-up occurs when snowline is closer to the star due to the higher surface density there.
Therefore, a 2D study that self-consistently includes pebble dynamics, vapor transport and phase change processes—while solving for the disk temperature and density structure—is required to assess the impact of the snowline on the characteristics of the solid pile-up.

A proper physical description of the snowline is also important for the post-planetesimal phases of planet formation.
After planetesimals are formed in the pile-up, a substantial enhancement in solids would enable fast growth of planets \citep{OrmelEtal2017, LiuEtal2019,SchoonenbergEtal2019}. 
\citet{OrmelEtal2017} proposes that snowline can act as a planetary ``factory'', producing planets sequentially to form compact planetary systems like the Trappist-1 \citep{GillonEtal2016,GillonEtal2017}.
More generally, \citet{JiangEtal2023} demonstrates that dust rings, where solids are trapped, can spawn planets even at tens of au, where the ubiquitous rings are usually observed by the Atacama Large Millimeter/submillimeter Array (ALMA) (e.g., \citealt{HuangEtal2018}). As the snowline delineates the solid and gas composition of the disk (e.g., \citealt{OebergEtal2023}), the density distributions of ice and vapor across it—such as the width, strength and composition of the pile-up—will determine the composition of planets that form at the snowline region \citep{SchoonenbergEtal2019,JohansenEtal2021}.

In summary, given all these important roles of snowline in planet formation, a realistic, 2D description of the snowline structure will enhance our understanding of how the snowline influences the formation of the first generation of planetsimals, the growth of planetary embryos into planets, and the composition of the planets that ultimately emerge. 
Besides, self-consistently solving the temperature and density structures could help identify potential observation imprints of water snowlines, which still remain elusive (e.g., \citealt{Zhang2024}). 

In this work, we perform 2D multifluid hydrodynamic simulations in the disk's radial ($r$)-vertical ($z$) plane with the inclusion of sublimation and condensation processes.
Pebbles are modeled as compounds containing various chemical components (e.g., water ice, silicates). Each component is treated as a single pressureless dust fluid, whose dynamics is tracked using the multifluid dust module in \texttt{Athena++} \citep{HuangBai2022}. In addition, vapor species are treated as passive scalars. The mass transfer and latent heat energy exchange during sublimation or condensation are self-consistently solved with the recently-developed phase change module \citep{WangEtal2023}. This novel setup enables simultaneous tracking of gas flow, pebble dynamics and the phase change between ice and vapor in hydrodynamic simulations. 
While we omit, for simplicity, dust collisional processes (coagulation or fragmentation), the pebble size can be altered by vapor recondensation and ice sublimation.
Furthermore, to determine the temperature structure near the snowline, we adopt a two-stream radiation transfer method (e.g., \citealt{MoriEtal2019}) that takes into account stellar irradiation, viscous heating, heat diffusion and latent heat exchange.

The paper is structured as follows. In \se{methods}, we present the 2D snowline model, covering the multifluid hydrodynamic equations for gas, vapor, and pebbles, phase change processes, and the radiation transfer method. \Se{results} begins with describing the typical 2D steady-state snowline structure for the active, passive, and vertically-isothermal disk setups. Then we present the impact of the snowline morphology on the solid pile-up, highlighting a "water-cycle" that promotes solid pile-up in active disks.
In \se{latent_heat} we examine the effect of latent heat, which varies in different types of snowline thermal structure. \Se{analysis} presents a detailed flow pattern analysis and Lagrangian water parcel tracking to understand the build-up of pile-up and the "water-cycle" phenomenon in 2D. In \se{discussion}, We discuss the implications on planetesimal formation, pebble accretion and the observational imprints left by latent heat effect. We further discuss the caveats and future improvements of this work. \Se{conclusion} lists the main conclusions.

\section{Methods}
\label{sec:methods}
This section describes the 2D hydrodynamic snowline model and presents the choices of simulation parameters. \se{disk_model} outlines the gas disk model. Then \se{hydro_eqs} presents the governing equations of the gas, pebble and vapor fluids, and the aerodynamic properties of pebbles are detailed in \se{peb_dynamics}. The treatments of ice sublimation and vapor condensation are described in \se{phase_change} and the radiation transfer method in \se{RT}. 
Finally, \se{simu_cases} presents the design of simulation cases explored in this work and summarize their  corresponding parameter space.

\subsection{Gas disk model}
\label{sec:disk_model}
We set up a 2D disk model in the radial ($r$) and vertical ($z$) direction in a Cylindrical coordinate.
Throughout the paper, the $\alpha$-disk model \citep{ShakuraSunyaev1973} is applied, where the viscosity is prescribed as,
\begin{equation}
    \nu = \alpha c_{s} H_{\mathrm{g}},
\end{equation}
where $c_{s} = \sqrt{k_{\mathrm{B}}T/\mu m_{\mathrm{p}}}$ is the local sound speed, $\mu$ is the mean molecular weight, $H_{\mathrm{g}} = c_{s}/ \Omega$ is the gas scale height and $\alpha$ is the nondimensional viscosity parameter. Assuming that the gas gets accreted to the central star at a steady rate $\dot{M}_{\mathrm{acc}}$, the gas surface density can be obtained as (e.g., \citealt{Armitage2020}),
\begin{equation}
    \label{eq:Sigma_g}
    \Sigma_{\mathrm{g}} = \frac{\dot{M}_{\mathrm{acc}}}{3\pi \nu}.
\end{equation}
In reality, \md{the} viscosity varies with temperature \md{and therefore} altitude in the disk. However, to simplify the disk profile, we fix the viscosity to be a power-law function of radius only, $\nu = \nu_{0} (r/r_{0})^{b}$, where $\nu_{0} = \alpha c_{s,0} H_{\mathrm{g},0}$ is the  viscosity at \md{the reference position} $r_{0}$. \md{We choose $r_{0}=3.0~\mathrm{au}$ and the power law index $b=-1$ following \citet{SchoonenbergOrmel2017}}. In this way, given $\dot{M}_{\mathrm{acc}}$ and $\alpha$, the surface density profile is fixed. 
The turbulent viscosity also causes diffusion to the gaseous components. We set the turbulent diffusivity $D_{\gas}$ of vapor equal to the viscosity $\nu$.

\subsection{Governing equations}
\label{sec:hydro_eqs}
The hydrodynamic equations of gas, pebble and vapor tracer fluids are presented in conservative form. 
The subscripts ``g'' and ``p'' are used to denote gas and pebble, respectively.
We refer to \citet{WangEtal2023} for detailed design of this hydyodynamic system. The governing equations read:

\begin{align}
\label{eq:CE_gas}
    &\frac{\partial \rho_{\gas}}{\partial t} + \nabla \cdot (\rho_{\gas} \bm{v}_{\gas}) = 0, \\
\label{eq:ME_gas}
    &\frac{\partial (\rho_{\gas} \bm{v}_{\gas})}{\partial t} + \nabla \cdot (\rho_{\gas} \bm{v}_{\gas} \bm{v}_{\gas} + P_{\gas} \bm{I} + \bm{\Pi}_{\nu}) = \rho_{\gas} \bm{f}_{\mathrm{g, src}}, \\
\label{eq:EE}
    &\frac{\partial E_{\gas}}{\partial t} +\nabla \cdot\left[\left(E_{\gas}+P_{\gas}\right) \bm{v}_{\gas} + \bm{\Pi}_{\nu} \cdot \bm{v}_{\gas} \right]  = \rho_{\gas} \bm{f}_{\gas, \mathrm{src}} \cdot \bm{v}_{\gas}, \\
\label{eq:CE_dust}
    & \frac{\partial \rho_{\parti, i}}{\partial t}+\nabla \cdot\left(\rho_{\parti, i} \bm{v}_{\parti}+\mathcal{F}_{\mathrm{p, dif}, i}\right)=0, \\
\label{eq:ME_dust}
    & \frac{\partial \rho_{\parti, i}\left(\bm{v}_{\parti}+\bm{v}_{\parti, \mathrm{dif}, i}\right)}{\partial t}+\nabla \cdot\left(\rho_{\parti, i} \bm{v}_{\parti} \bm{v}_{\parti}+\bm{\Pi}_{\mathrm{dif}, i}\right) = \nonumber \\
    & \rho_{\parti, i} \bm{f}_{\parti, \mathrm{src}, i}+\rho_{\parti, i} \frac{\bm{v}_{\mathrm{g}}-\bm{v}_{\parti}}{t_{\mathrm{s}}}, \\
\label{eq:CE_vapor}
    & \frac{\partial \rho_{\mathrm{vap}}}{\partial t}+\nabla \cdot\left(\rho_{\mathrm{vap}} \bm{v}_{\gas}+\mathcal{F}_{\mathrm{dif, vap}}\right)=0.
\end{align}
where $\bm{v}_{\gas}$ and $\bm{v}_{\parti}$ represent the velocities of gas and pebbles, respectively. Tracer fluid (passive scalar) has the same velocity as gas. The tensors $\bm{\Pi}_{\nu}$ and $\bm{\Pi}_{\mathrm{dif},i}$ denote the viscous stress of gas and diffusion 
 of momentum flux respectively. For each component $i$ (ice or silicate) of the pebble, the particle concentration flux is given by,
\begin{equation}
    \mathcal{F}_{\mathrm{p, dif}, i} = -\rho_{\gas} D_{\parti} \nabla \left(\frac{\rho_{\parti,i}}{\rho_{\gas}} \right) \equiv \rho_{\parti,i} \bm{v}_{\parti, \mathrm{dif}, i},
\end{equation}
which defines the effective diffusive velocity $\bm{v}_{\parti, \mathrm{dif}, i}$. The last term in \eq{ME_dust} represents the aerodynamic drag experienced by the particles, which is controlled by the stopping time $t_{\mathrm{s}}$. The backreaction of solids acting on gas is neglected in this work (see \se{pltsml_formation}). The phase change module tracks the vapor fraction using passive scalars, as shown in \eq{CE_vapor}. Finally, $\bm{f}_{\mathrm{src}}$ represents the source term due to external forces, which only includes stellar gravity here.

The hydrodynamic equations are solved in a spherical-polar coordinate with \texttt{Athena++} \citep{StoneEtal2020}, where the pebble dynamics is computed by the multifluid dust module \citep{HuangBai2022}.

\subsection{Pebble dynamics}
\label{sec:peb_dynamics}
Due to the aerodynamic drag of the gaseous disk, pebbles loss angular momentum go through inward radial drift. 
The aerodynamic properties of pebbles can be characterized by the stopping time ($t_{\mathrm{s}}$). To obtain $t_{\mathrm{s}}$, we account for two drag regimes: Epstein drag and Stokes drag. Then the stopping time is given by,
\begin{equation}
    t_{\mathrm{s}}=\left\{
    \begin{aligned}
    \frac{\rho_{\bullet, \parti} s_{\parti}}{v_{\mathrm{th}} \rho_{\gas}} & , & (\mathrm{Epstein:}  \quad s_{\mathrm{p}}<\frac{9}{4} l_{\mathrm{mfp}}  ), \\
    \frac{4 \rho_{\bullet, \parti} s_{\parti}^{2}}{9 v_{\mathrm{th}} \rho_{\gas}l_{\mathrm{mfp}}} & , & (\mathrm{Stokes:}  \quad s_{\mathrm{p}}>\frac{9}{4} l_{\mathrm{mfp}}  ),
    \end{aligned}
    \right.
\end{equation}
where the mean free path of the gas molecules $l_{\mathrm{mfp}}$,
\begin{equation}
    l_{\mathrm{mfp}} = \frac{\mu m_{\mathrm{p}}}{\sqrt{2} \rho_{\mathrm{g}} \sigma_{\mathrm{mol}}},
\end{equation}
with $\sigma_{\mathrm{mol}}$ is the molecular collision cross-section, which we approximate by the collision cross-section of molecular hydrogen $\sigma_{\mathrm{mol}}=2 \times 10^{-15} \mathrm{~cm}^{2}$ \citep{ChapmanCowling1991}, $s_{\parti}$ is the pebble size, $\rho_{\bullet, \parti}$ is the pebble internal density and $v_{\mathrm{th}} = \sqrt{8/\pi}c_{\mathrm{s}}$ is the thermal velocity of the gas molecules.

We assume that pebbles \md{drift} towards the water snowline \md{with a steady mass flux and} consist of two material components: refractory silicate and volatile water ice. The water ice mass fraction of the particles, $\zeta = \rho_{\parti,\mathrm{ice}} / (\rho_{\parti,\mathrm{ice}} + \rho_{\parti,\mathrm{sil}})$ is assumed $\zeta = 0.5$ for pebbles that enter the simulation domain.
\md{Therefore, pebbles are injected at a rate following $\dot{M}_{\mathrm{peb}} = 2f_{\mathrm{i/g}} \dot{M}_{\mathrm{acc}}$, where $f_{\mathrm{i/g}}$ is the ice-to-gas flux ratio.}

However, $\zeta$ is allowed to vary in the simulation due to sublimation and condensation processes.
The internal density $\rho_{\bullet, \parti}$ of a pebble compound \md{changes accordingly},
\begin{equation}
    \rho_{\bullet, \parti} = \left(\frac{1-\zeta}{\rho_{\bullet,\mathrm{sil}}} + \frac{\zeta}{\rho_{\bullet,\mathrm{ice}}} \right)^{-1},
\end{equation}
where $\rho_{\bullet, \parti}$ is the internal density of pebble. We adopt $\rho_{\bullet, \mathrm{sil}} = 3.0~\mathrm{g}\mathrm{~cm}^{-3}$ for pure silicate and $\rho_{\bullet, \mathrm{ice}} = 1.0~\mathrm{g}\mathrm{~cm}^{-3}$ for pure water ice.

In order to calculate the stopping time, we need to determine the pebble size $s_{\parti}$. We consider the internal structure of pebble such that ice is coated on top of the silicates (``single-seed model'' in the terminology of \citealt{SchoonenbergOrmel2017}). When pebbles cross the snowline and ice sublimates, it is assumed that the bare silicate core remains intact (\citealt{SpadacciaEtal2022,HougeEtal2024}, and see \citealt{SaitoSirono2011, HyodoEtal2019} for case where pebbles disaggregate). Since the mass of the silicate core ($m_{\mathrm{core}}$) in each pebble is unaffected by ice sublimation, the pebble number density can be obtained as,
\begin{equation}
    n_{\mathrm{p}} = \frac{\rho_{\mathrm{p,sil}}}{m_{\mathrm{core}}},
\end{equation}
where $\rho_{\mathrm{p,sil}}$ is the volume density of silicate. From the number density, the particle mass is obtained, $(\rho_{\mathrm{p,sil}} + \rho_{\mathrm{p,ice}})/n_\parti$,  which, combined with its internal density, gives $s_{\parti}$:
\begin{equation}
\label{eq:peb_size}
    \frac{4\pi}{3} s_{\parti}^3 = \frac{1}{\rho_{\bullet, \parti}} \frac{ \rho_{\mathrm{p,sil}} + \rho_{\mathrm{p,ice}}}{n_{\parti}}
    = \frac{m_\mathrm{core}}{\rho_\mathrm{\bullet,sil}} \left( 1 + \frac{\rho_\mathrm{p,ice}}{\rho_\mathrm{p,sil}}\frac{\rho_\mathrm{\bullet,sil}}{\rho_\mathrm{\bullet,ice}} \right).
\end{equation}
In summary, the stopping time of the pebble is jointly determined by the densities of silicate and ice in the pebble compound. In the multifluid dust module, at each grid cell, the silicate and ice fluids share the same stopping time.

Apart from drift, pebbles also diffuse. Following \citet{YoudinLithwick2007}, the diffusivity of solids is related to the gas diffusivity through the Schmidt number $\mathrm{Sc}$,
\begin{equation}
    D_{\parti} = D_{\gas} \times \mathrm{Sc} = \frac{D_{\gas}}{1 + \tau_{\mathrm{s}}^{2}},
\end{equation}
where $\tau_{\mathrm{s}} = \Omega t_{\mathrm{s}}$ is the dimensionless stopping time (Stokes number).

\md{Throughout this work, we fix $f_{\mathrm{i/g}} = 0.4$, which corresponds to a pebble-to-gas flux ratio of 0.8. Realistic disk evolution models invoking pebble growth suggest that $f_{\mathrm{i/g}}$ varies with e.g., disk metallicity, disk size and time \citep{BirnstielEtal2010,BirnstielEtal2012,LambrechtsJohansen2014,IdaEtal2016,SchoonenbergEtal2018}. As shown in \citet{SchoonenbergEtal2018}, the pebble-to-gas flux ratio can readily reach unity at early times (${<}10^{5}$ yr) of the disk under solar metallicity. At later times, the flux ratio may drop a bit \citep{IdaEtal2016} but we do not expect this to qualitatively affect our results. The initial pebble size is set such that the Stokes number $\tau_{\mathrm{s}}$ is  $\tau_{\mathrm{s,0}}$ at $r_{0}$. In all runs, $\tau_{\mathrm{s,0}} = 0.03$ is used, following \citet{SchoonenbergOrmel2017}, which corresponds to a particle radius of $s_{\parti} \approx 2~\mathrm{cm}$ in our disk model. With ice sublimation and vapor recondensation, the pebble size can decrease and increase following \eq{peb_size}. We do not include coagulation processes in our model.}

\subsection{Phase change}
\label{sec:phase_change}
When pebbles drift across the snowline, the water ice component sublimates. This process transfers mass from ice to vapor, but also absorbs latent heat from the environment. Furthermore, the released water vapor, which has a higher mean molecular weight, will alter the pressure support thus the pressure scaleheight of the disk.

To properly capture the hydrodynamic and thermodynamic effect of pebble sublimation, we apply the phase change module developed in \citet{WangEtal2023}. This module treats mass transfer and energy exchange during ice sublimation and vapor condensation self-consistently. The water vapor is treated as an additional tracer fluid, which advects with the gas flow and diffuses according to the diffusivity $D_{\gas}$.

\md{Sublimation (condensation) occurs following rate expression (e.g., \citealt{RosJohansen2013, SchoonenbergOrmel2017}). Within one timestep ($\Delta t$), the amount of vapor gained during sublimation $\left(\Delta \rho_{\mathrm{vap}}\right)_{\rm subl}$ or ice gained during condensation $\left(\Delta \rho_{\mathrm{ice}}\right)_{\rm cond}$ is, }
\begin{equation}
\label{eq:limiter}
     \left(\Delta \rho_{\mathrm{vap}}\right)_{\rm subl} = \left(\Delta \rho_{\mathrm{p,ice}}\right)_{\rm cond} = \pi s^{2}_{\parti}v_{\mathrm{th,vap}} \rho_{\mathrm{vap}} n_{\parti}  \left|1 - \frac{P_{\mathrm{eq}}}{P_{\mathrm{vap}}}\right| \cdot \Delta t,
\end{equation}
\md{where $n_\mathrm{p}$ is the pebble number density, $v_{\mathrm{th,vap}}$ is the thermal velocity of vapor, $P_{\mathrm{vap}}$ is the vapor partial pressure, and $P_{\mathrm{eq}}$ is the saturation pressure of vapor as given by the \textit{Clausius-Clapeyron} equation:}
\begin{equation}
    \label{eq:P_satur}
    P_{\mathrm{eq}} = P_{\mathrm{eq,0}} \exp (-T_{a}/T).
\end{equation}
\md{Here $P_{\mathrm{eq,0}}$ and $T_{a}$ are constants specific to the species. For water, we take $T_{a} = 6062~\mathrm{K}$ and $P_{\mathrm{eq,0}}=1.14\times 10^{13}~\mathrm{g~cm^{-1}s^{-2}}$ \citep{LichteneggerKomle1991}.} 
\md{However, sublimation (condensation) stalls when the vapor equilibrium state given by \eq{P_satur} is reached. Therefore, in the phase change module, the amount of material that is transferred between the phases is subject to the condition that vapor equilibrium is obtained whenever possible, i.e.,}
\begin{equation}
\label{eq:phase_change_equil}
\begin{aligned}
    \Delta \rho_{\mathrm{vap}} &= \min \left[\left(\Delta \rho_{\mathrm{vap}}\right)_{\rm subl},\rho_{\mathrm{eq}} - \rho_{\mathrm{vap}} \right];\\
    \Delta \rho_{\mathrm{ice}} &= \min\left[\left(\Delta \rho_{\mathrm{ice}}\right)_{\rm cond},\rho_{\mathrm{vap}} - \rho_{\mathrm{eq}} \right],
\end{aligned}
\end{equation}
\md{where $\rho_{\mathrm{eq}}$ is the vapor density corresponding to the saturation pressure.}  \md{In this way, at the disk midplane around the snowline, vapor tends to follow the saturation pressure, while at the disk atmosphere, vapor tends to be super-saturated due to the slow condensation rate (\eq{limiter}) and can be freely transported by diffusion and advection.}

Latent heat is defined as the enthalpy difference between the vapor and the ice. Supposing that $\delta \rho$ is the mass transfer from ice to vapor during sublimation, an additional energy $\delta E = L_{\mathrm{heat}} \delta \rho$ should be subtracted from the total internal energy, while the same amount of energy will be released if condensation occurs. For water ice, we take $L_{\mathrm{heat}} = 2.75 \times 10^{10}~\mathrm{erg~g^{-1}}$ \citep{FraySchmitt2009}. For the description of the phase change module and its detailed implementation in \texttt{Athena++}, we refer the reader to \citet{WangEtal2023}. 

The ideal equation of state of gas is adopted in this work,
\begin{equation}
    P_{\gas} = \rho_{\gas} \frac{k_{\mathrm{B}} T}{\mu m_{\mathrm{p}}}
\end{equation}
Once vapor is released, we change the mean molecular weight of the gas accordingly following,
\begin{equation}
\label{eq:mmw}
    \frac{1}{\mu} = \frac{f_{\mathrm{v}}}{\mu_{\mathrm{H_{2}O}}} + \frac{1 - f_{\mathrm{v}}}{\mu_{\mathrm{xy}}},
\end{equation}
where $f_{\mathrm{v}}$ is the vapor mass fraction, and $\mu_{\mathrm{H_{2}O}} = 18$ and $\mu_{\mathrm{xy}} = 2.34$ are the mean molecular weight of water and H-He mixture respectively.

\subsection{Two-stream radiation transfer}
\label{sec:RT}
In this work, we focus on the steady state solution of the snowline region rather than dynamical and time-dependent behaviors. Thus, we opt to determine the temperature structure using an equilibrium solution obtained with the dissipation profile.
Following \cite{MoriEtal2019}, we define the total dissipation profile as,
\begin{equation}
    q_{z} = q_{\mathrm{irr}} + q_{\mathrm{vis}} + q_{\mathrm{latent}} + q_{\mathrm{diff}},
\end{equation}
where $q_{\mathrm{irr}}$, $q_{\mathrm{vis}}$, $q_{\mathrm{latent}}$ and $q_{\mathrm{diff}}$ are heating terms due to stellar irradiation, viscous dissipation, latent heat exchange and heat diffusion, respectively.
Here we implicitly adopt the two-stream assumption, where the incoming stellar irradiation and disk thermal emission are well separated in the visible and infrared band respectively so that we can combine the corresponding heating sources independently.

Following \citet{CalvetEtal1991},
the heating rate contributed by stellar irradiation $q_{\mathrm{irr}}$ is,
\begin{equation}
    q_{\mathrm{irr}} = E_{0} \rho \kappa_{\mathrm{vi}} \left[ \exp \left( -\frac{\tau_{\mathrm{vi}(z)}}{\mu_{0}} \right) + \exp \left(- \frac{\tau_{\mathrm{vi}}(-\infty) - \tau_{\mathrm{vi}}(z) }{\mu_{0}} \right) \right].
\end{equation}
Here $E_{0} = L_{\star}/(8\pi r^{2})$ is the stellar irradiation flux and $\kappa_{\mathrm{vi}}$, $\tau_{\mathrm{vi}}$ are the opacity and optical depth for visible light respectively:
\begin{equation}
    \tau_{\mathrm{vi}} = \int_{z}^{+\infty} \rho \kappa_{\mathrm{vi}} \mathrm{d} z^{\prime}.
\end{equation}
Also, $\mu_{0}$ is the cosine of the incident angle of the stellar irradiation,
\begin{equation}
    \mu_{0} = r \frac{\mathrm{d}}{\mathrm{d}r}\left( \frac{H_{\mathrm{p}}}{r}\right),
\end{equation}
and $H_{\mathrm{p}}$ is the height of the photosphere of the disk \citep{ChiangGoldreich1997}. We take $H_{\mathrm{p}} = 4 H_{\mathrm{g}}$ in our modelling. In principle, ray tracing is needed to calculate the irradiative heating accurately. However, this simplification will not affect the overall results since the ice-sublimating regions are always well within optically thick limit for stellar irradiation in all runs.

For active disks, where viscous dissipation is deposited in the midplane, the viscous heating term $q_{\mathrm{vis}}$ (e.g., \citealt{Armitage2020}) follows,
\begin{equation}
    q_{\mathrm{vis}} = \frac{9}{4} \rho \nu \Omega^{2}.
\end{equation}
To include the thermal effect of ice sublimation (condensation), we model the latent heat absorption (release) as an additional heating term defined as,
\begin{equation}
    q_{\mathrm{latent}} = -\frac{\mathrm{d} \rho_{\mathrm{vap}}}{\mathrm{d} t} \cdot L_{\mathrm{latent}}.
\end{equation}
The \md{change of vapor mass} $\mathrm{d} \rho_{\mathrm{vap}} / \mathrm{d} t$ is measured in the simulation, as determined by the phase change module.

Lastly, in optically thick regions, unlike optically thin regions where vertical transfer is usually quicker than the radial component due to the small aspect ratio of the disk, heat diffusion works both vertically and radially. We aim to complete our radiative transfer approach by including the radial heat diffusion only in the optically thick region. Therefore, following \citet{XuKunz2021},
\begin{equation}
    q_{\mathrm{diff}} = \nabla\cdot \left[\kappa_{\mathrm{heat}} \nabla T \exp \left( -1/\tau_{\mathrm{R}} \right) \right],
\end{equation}
and,
\begin{equation}
\kappa_{\mathrm{heat}} = \frac{16 \sigma_{\mathrm{SB}} T^{3}}{3 \kappa_{\mathrm{R}}\rho}.
\end{equation}
Here $\kappa_{\mathrm{heat}}$ is the heat conduction coefficient evaluating the efficiency of energy transport by radiative diffusion (e.g., \citealt{KippenhahnEtal2013}), $\kappa_{\mathrm{R}}$ the Rosseland mean opacity and $\tau_{R}$ the corresponding optical depth. The exponentially diminishing term $\exp \left( -1/\tau_{\mathrm{R}} \right)$ suppresses the heat diffusion in optically thin regions.

Given the total dissipation profile $q_{z}$, the equilibrium solution of the temperature can be expressed as,
\begin{equation}
    \label{eq:T_z}
    T_{\mathrm{eq}}(z) = T_{\mathrm{eff}} \left( \frac{3}{4} \tau_{\mathrm{eff}} + \frac{\sqrt{3}}{4} + \frac{q_{z}}{4 \rho \kappa_{R} \mathcal{F}_{+\infty}} \right)^{1/4}.
\end{equation}
We refer the detailed derivation of \eq{T_z} to previous literature (e.g., \citealt{Hubeny1990,MoriEtal2019}) and briefly describe the meaning of the expression. First, $\mathcal{F}_{+\infty}$ is the radiative flux leavings the upper surface and $T_{\mathrm{eff}}$ is the corresponding effective temperature:
\begin{equation}
    \mathcal{F}_{+\infty} = \int_{0}^{+\infty} q_{z} \mathrm{d} z; \qquad T_{\mathrm{eff}} = (F_{+\infty} / \sigma_{\mathrm{SB}})^{1/4}.
\end{equation}
Further, $\tau_{\mathrm{eff}}$ is the flux-weighted effective optical depth,
\begin{equation}
    \tau_{\mathrm{eff}}(z) = \frac{1}{\mathcal{F}_{+\infty}} \int_{z}^{+\infty} \rho \kappa_{\mathrm{R}} \mathcal{F}(z^{\prime}) dz^{\prime},
\end{equation}
where $\mathcal{F}(z) = \int_{0}^{z} q_{z} dz$. If a high amount of heating is deposited in optically thick regions, $\tau_{\mathrm{eff}}$ will dominated \eq{T_z}. In contrast, in the optically thin regions, the temperature is determined by a local balance of heating and cooling terms  (Eq.\ref{eq:T_z}, third term) and the radiation field sets the background equilibrium temperature (Eq.\ref{eq:T_z}, second term).

In the simulation, in order to stabilize the system, the temperature is gradually relaxed  by introducing a thermal relaxation parameter $\beta$:
\begin{equation}
    \frac{\mathrm{d} T}{\mathrm{d}t} = \frac{T_{\mathrm{eq}} - T_{0}}{\beta \Omega^{-1}},
\end{equation}
where $T_{0}$ is the temperature before the thermal relaxation. Although this relaxation breaks the equilibrium solution obtained with the two-stream radiation transfer method, the steady state solution will remain consistent, since $\mathrm{d} T / \mathrm{d} t \approx 0$. We choose $\beta = 50$ in our simulation to prevent numerical oscillations.

\subsection{Simulation cases}
\label{sec:simu_cases}
Our aim is to investigate the impact of qualitatively distinct disk thermal structures on the characteristics of the snowline region. Consequently, we define three disks:
\begin{itemize}
    \item \underline{Passive disks}. The temperature is determined by how deep the stellar irradiation ($q_{\mathrm{irr}}$) penetrates while the contribution from viscous heating is negligible. The disk is optically thick for cooling. Passive disks feature a hot upper layer and a cold midplane. 
    \item \underline{Active Disks}. Here the viscous heating ($q_{\mathrm{vis}}$) dominates, which, combined with the optically thick environment that prevents cooling, renders the temperature higher near the midplane.
    \item \underline{Isothermal Disks}. The main heating source is also the stellar irradiation as the passive disks. However, the disk is optically thin for cooling.
\end{itemize}
These three cases represent the main snowline thermal morphology investigated in this work. 

Because we aim to directly compare the simulations to previous 1D calculations \citep{DrazkowskaAlibert2017,SchoonenbergOrmel2017, HyodoEtal2021}, we set up the disk model by fixing $\alpha = 3\times 10^{-3}$ and $\dot{M}_{\mathrm{acc}} = 10^{-8} M_{\odot} ~\mathrm{yr}^{-1}$ , following \citet{SchoonenbergOrmel2017}. In addition, we anchor the midplane snowline location ($r_{\mathrm{snow,mid}}$) at ${\approx}2.1$ au before the injection of icy pebbles (which will shift $r_{\mathrm{snow}}$ when sublimating) by tuning the stellar luminosity and opacity (see below). The location of the snowline is approximated by equating the vapor saturation pressure to the expected vapor pressure if all incoming ice is sublimated \citep{SchoonenbergOrmel2017}.
\begin{equation}
    \label{eq:r_snow}
    P_{\mathrm{eq,0}} \exp \left[-\frac{T_{a}}{T}\right] = \frac{k_{B} T}{\mu} \rho_{\mathrm{g, mid}} f_{\mathrm{i/g}},
\end{equation}
where $\rho_{\mathrm{g, mid}}$ is the midplane gas density. In practice, given a disk viscosity profile and mass flux, the surface density profile can be obtained. Fixing the surface density at a certain radius, the gravity-balanced vertical density and temperature distribution can be iteratively solved. This serves as the initial condition of the simulation and supplies the initial estimate of $r_{\mathrm{snow}}$.

Under this simplified disk model, the main parameters that determine the initial thermal structure are $\kappa_{\mathrm{vi}}$, $\kappa_{\mathrm{R}}$ and $L_{\star}$, which are all highly uncertain among different disk environments (e.g., \citealt{BirnstielEtal2018}). To simplify, we use constant Rosseland-mean opacity with respect to the density of non-condensable gas (H and He) $\tau_{\mathrm{R}} = \int \kappa_{\mathrm{R}} \rho_{\mathrm{H,He}} \mathrm{d} z$ (\tb{simu_cases}). Also, we use the same visual band opacity $\kappa_{\mathrm{vi}} = 10 ~ \mathrm{cm^{2}g^{-1}}$ among all runs. This value corresponds approximately to the opacity of a grain size distribution with the maximum grain size ($a_{\mathrm{max}}$) around sub-mm and a power-law of $-3.5$ \citep{MathisEtal1977}, as calculated from the DSHARP opacity model \citep{BirnstielEtal2018}.

By varying $\kappa_{\mathrm{R}}$ and $L_{\star}$, we construct disks with qualitatively-distinct thermal morphology and their initial midplane snowlines at 2.1 au. As shown in \fg{r_snow_init}, the permitted parameter space can be divided into three regimes: ``passive disk'' when $L_{\star}$ is high, ``active disk'' when $\kappa_{\mathrm{R}}$ is large and ``vertically-isothermal'' disk when $\kappa_{\mathrm{R}}$ is small. The chosen parameters of simulation runs are denoted by black dots in \fg{r_snow_init} and summarized in \tb{simu_cases}.

As test cases to validate our model, we also conduct 1D snowline simulations (\tb{simu_cases}). \md{The \texttt{1D/SO17} run is comparable to the ``simple, single-seed'' model in \citet{SchoonenbergOrmel2017} except that the increase of mean molecular weight with vapor injection is included here. The \texttt{1D-hydro} run has the gas motion solved hydrodynamically, yielding distinct results, which is discussed in \app{1D_model} and \se{pltsml_formation}. }

\begin{table*}[t]
    \centering
    \caption{Summary of simulation runs with their selected parameters and \md{characteristic outputs}.}
    \label{tab:simu_cases}
    \begin{tabular}{lcccccccccc} 
        \hline
        Case & $\kappa_{\mathrm{R}}$ & $L_{\star}$ & $q_{\mathrm{latent}}$ & $r_\mathrm{snow,mid}$ & $\xi_{\mathrm{pk}}$ & $r_{\mathrm{pk}}$ & FWHM & $M_{\mathrm{ice}}$ &  Fig. \\
        Unit & (cm$^{2}$ g$^{-1}$) & ($L_{\odot}$) & - &  (au) & - & (au) & (au) & ($M_{\oplus}$) & - \\
        \hline
        \texttt{active} & 4.0 & 1.0 & Y & 1.70 & 0.53 & 2.04 & 0.69 & 14.06 & \ref{fig:v_r},\ref{fig:latent_contra}\\
        \texttt{active-nqL} & 4.0 & 1.0 & N & 2.06 & 0.84 & 2.21 & 0.33 & 14.83 & \ref{fig:v_r},\ref{fig:latent_contra},\ref{fig:trace_contour}\\
        \texttt{passive} & 0.01 & 22.0 & Y & 2.02 & 0.62 & 2.32 & 0.50 & 10.70 & \ref{fig:profile_iso_highL},\ref{fig:v_r} \\
        \texttt{passive-nqL} & 0.01 & 22.0 & N & 2.05 & 0.66 & 2.32 & 0.43 & 10.71 & \ref{fig:trace_contour}\\
        \texttt{iso} & $1.2 \times 10^{-5}$  & 1.0 & Y & 1.91 & 0.53 & 2.38 & 0.62 & 11.31 & \\
        \texttt{iso-nqL} & $1.2 \times 10^{-5}$  & 1.0 & N & 2.06 & 0.65 & 2.33 & 0.44 & 10.59 & \\
        \texttt{1D/SO17} & $T(r)$ \textit{fixed} & - &  N & 2.02 & 0.28 & 2.24 & 1.06 & 4.02 & \ref{fig:1D_model},\ref{fig:output_summary} \\
        \texttt{1D-hydro} & $T(r)$ \textit{fixed} & -&  N & 2.02 & 0.48 & 2.23 & 0.42 & 8.18 & \ref{fig:1D_model},\ref{fig:output_summary} \\
        \hline
    \end{tabular}
    \tablefoot{In all runs, the ice-to-gas flux ratio $f_{\mathrm{i/g}}$ is fixed as 0.4 and the initial Stokes number of pebbles at the reference position $r_{0}$ is set to $\tau_{\mathrm{s,0}} = 0.03$. 1D runs labeled with the prefix ``\texttt{1D}'' have their temperature profiles fixed (\app{1D_model}). \md{The first 3 columns denote the simulation parameters (\se{simu_cases}) with $\kappa_{\mathrm{R}}$: the Rosseland-mean opacity,  $L_{\star}$: stellar luminosity and $q_{\mathrm{latent}}$: whether latent heat is included, with ``Y'' meaning it is and ``N'' indicating it is not. The following 5 columns denote the characteristic outputs of the simulation runs (\se{2D_structure}) with $r_\mathrm{snow,mid}$: the midplane snowline position,  $\xi_{\mathrm{pk}}$: the peak solid-to-gas ratio of the pile-up, $r_{\mathrm{pk}}$: the position of $\xi_{\mathrm{pk}}$, FWHM: the full-width-half maximum of the pile-up and $M_{\mathrm{ice}}$: the total ice mass stored in the pile-up. The last column refers to the figures where the runs are discussed. Figs. \ref{fig:contourf_density},\ref{fig:contourf_Tem},\ref{fig:eta} and \ref{fig:output_summary} present galleries of all 2D runs so they are not repeated in the column.}}
\end{table*}

\begin{figure}[t]
    \includegraphics[width=\columnwidth]{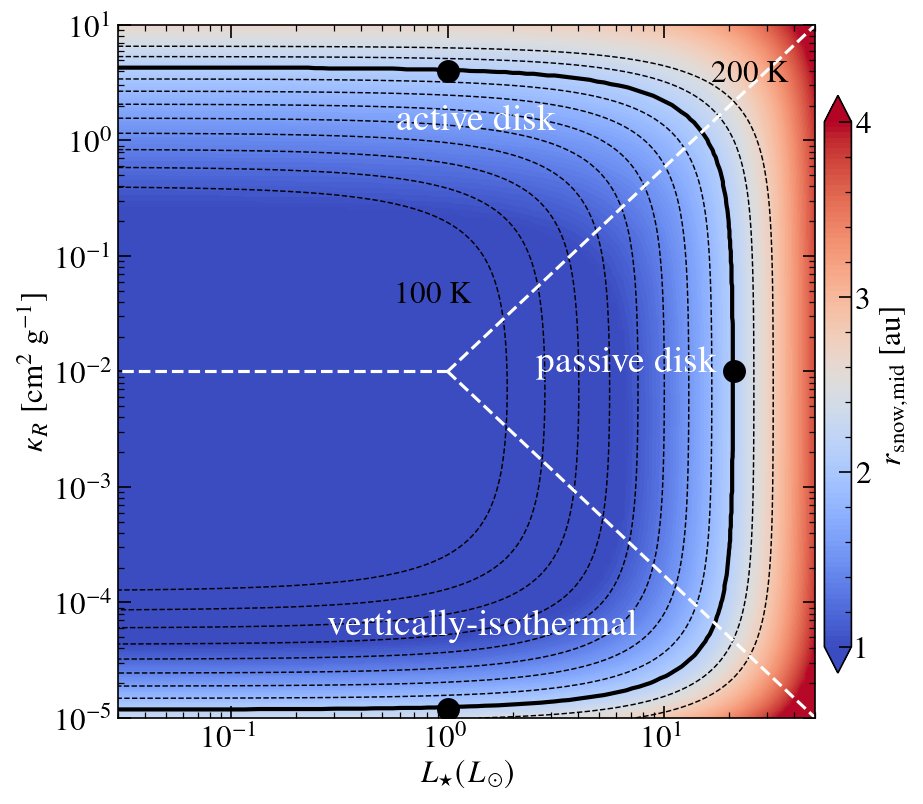}
    \caption{Initial midplane snowline location resulting from the two-stream radiation transport model (\eq{T_z}) as function of Rosseland mean opacity $\kappa_{\mathrm{R}}$ and stellar luminosity $L_{\star}$. Here the optical opacity is fixed to $\kappa_{\mathrm{vi}} = 10 ~ \mathrm{cm^{2}g^{-1}}$. The permitted parameter space is divided into three regimes (white dashed lines, for illustration purpose only): ``passive disk'' when $L_{\star}$ is high, ``active disk'' when $\kappa_{\mathrm{R}}$ is high and ``vertically-isothermal'' disk when $\kappa_{\mathrm{R}}$ is small. The dotted lines are temperature contours of the midplane at 2.1 au, ranging from 100 K to 200 K in 10 K increments. The black solid line represents the parameter combinations that lead to $r_{\mathrm{snow,mid}} = 2.1~\mathrm{au}$ and the black dots denote the parameter combinations used in this work, as summarized in \tb{simu_cases}. 
    Hydrodynamic and thermodynamic effects will \md{move} the snowline location in the simulation.}
    \label{fig:r_snow_init}
\end{figure}

\section{Results}
\label{sec:results}
In this section, we present the results of the 2D disk snowline models as listed in \tb{simu_cases} and introduced in \se{simu_cases}. We first present overall simulation setup and demonstrate that the simulations reach steady state based on the measured mass fluxes (\se{setup}). Then we introduce the temperature structure, density structure and water delivery patterns (\se{2D_structure}). In  \se{morphology}, we illustrate the effects of different snowline morphologies (active, passive, isothermal) have on the pile-up of solids at the snowline. We identify a ``water cycle'' that enhance the pile-up of water in the snowline region in active disks. The effects of latent heat will be discussed in \se{latent_heat}.

\begin{figure*}
	\centering
	\includegraphics[width=\textwidth]{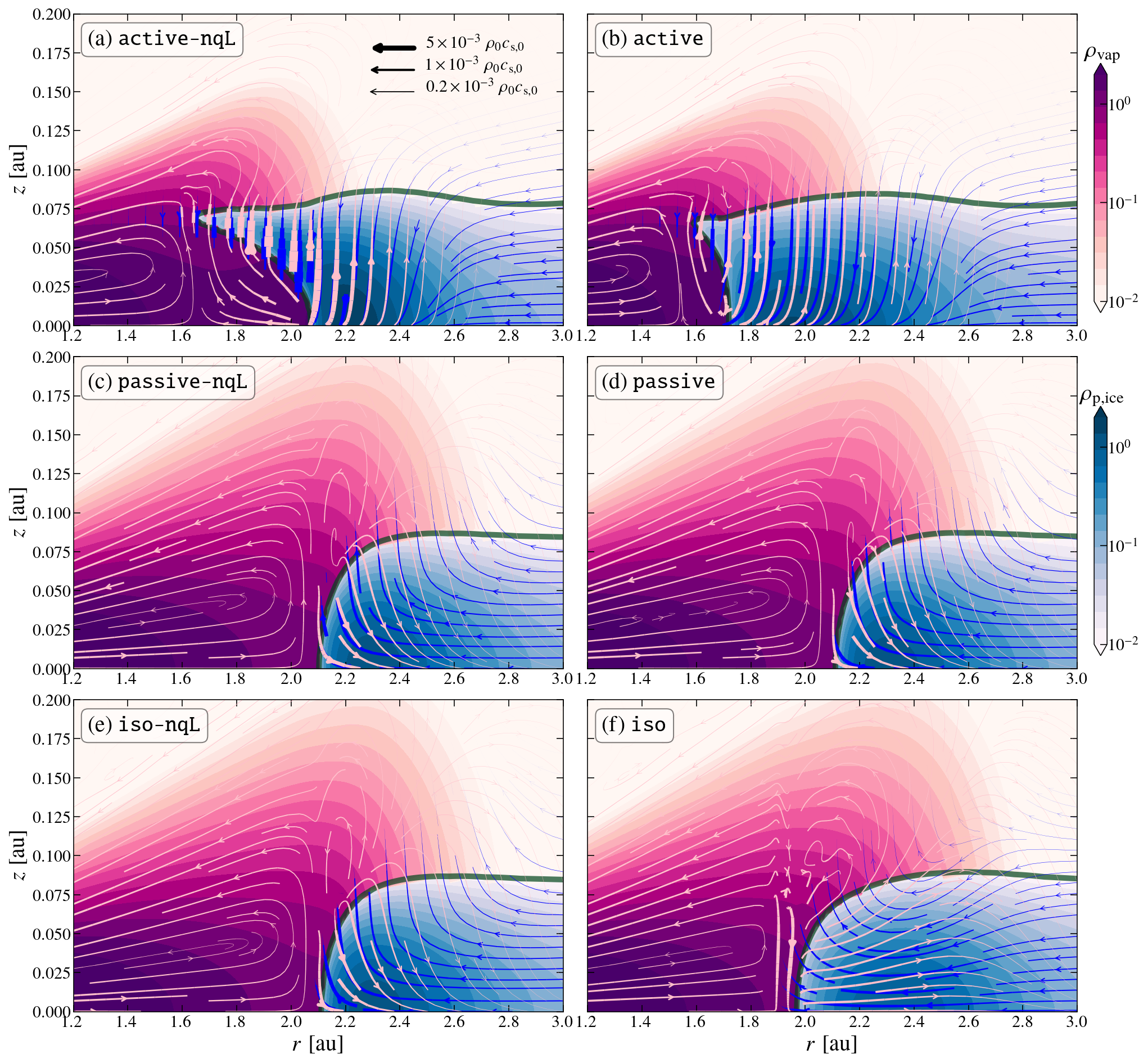}
	\caption{Steady state density structure of all simulated disk (as listed in \tb{simu_cases}). The dark green line denotes the boundary where $\rho_{\mathrm{p,ice}}/\rho_{\gas} = 10^{-2}$. The region enclosed by the green line filled with blue shading represents the ice density normalized to $\rho_{0} \approx 3.4\times10^{-11}~\mathrm{cm~g^{-3}}$ (the midplane gas density at $r_{0}$ of the unperturbed disk), while the region interior to the snowline indicates the vapor density with red shading. The ice (blue) and vapor (pink) mass fluxes are plotted with streamlines, whose thickness denotes the magnitude of the flux normalized to $\rho_{0} c_{\mathrm{s,0}}$. \md{The mass flows in the vertical direction, which are especially vigorous in the \texttt{active-nqL} and \texttt{active} runs, contribute towards an increased pile-up of ice in the snowline region.}
    }
	\label{fig:contourf_density}
\end{figure*}

\begin{figure*}
	\centering
	\includegraphics[width=1.95\columnwidth]{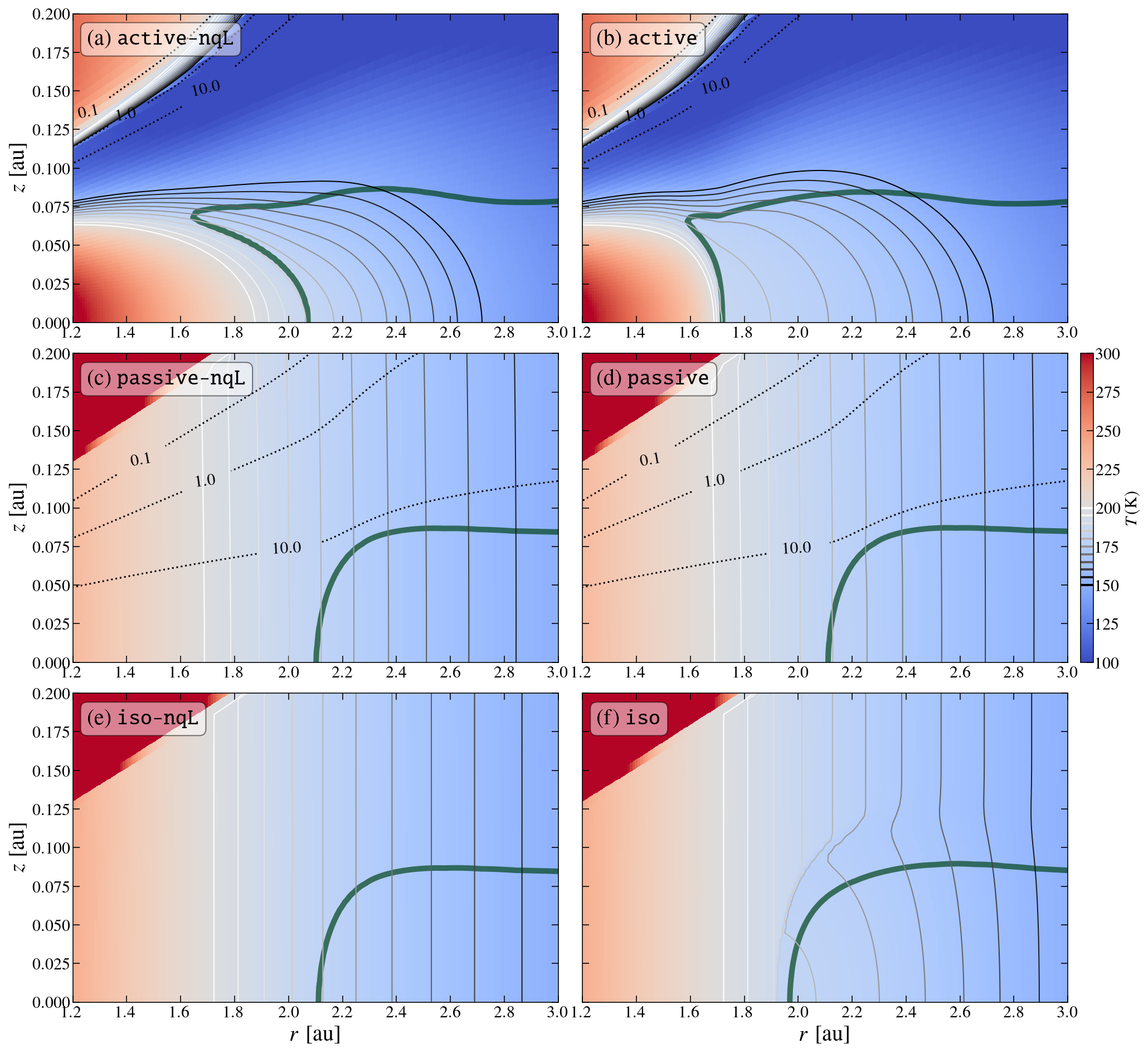}
	\caption{Steady state temperature structure of all the simulation \md{runs} (\tb{simu_cases}). The color shows the temperature structure with contours highlighting the range between 150 and 200 K \md{in increments of} 5 K. The dashed lines show the Rosseland mean optical depth $\tau_{\mathrm{R}}$ from $0.1$ to $10.0$, \md{displaying the} optically-thick or optically-thin \md{nature of the} cooling. In panel (e) and (f), the dashed lines are absent since the disk is very optically-thin.}
	\label{fig:contourf_Tem}
\end{figure*}

\begin{figure}
	\includegraphics[width=0.98\columnwidth]{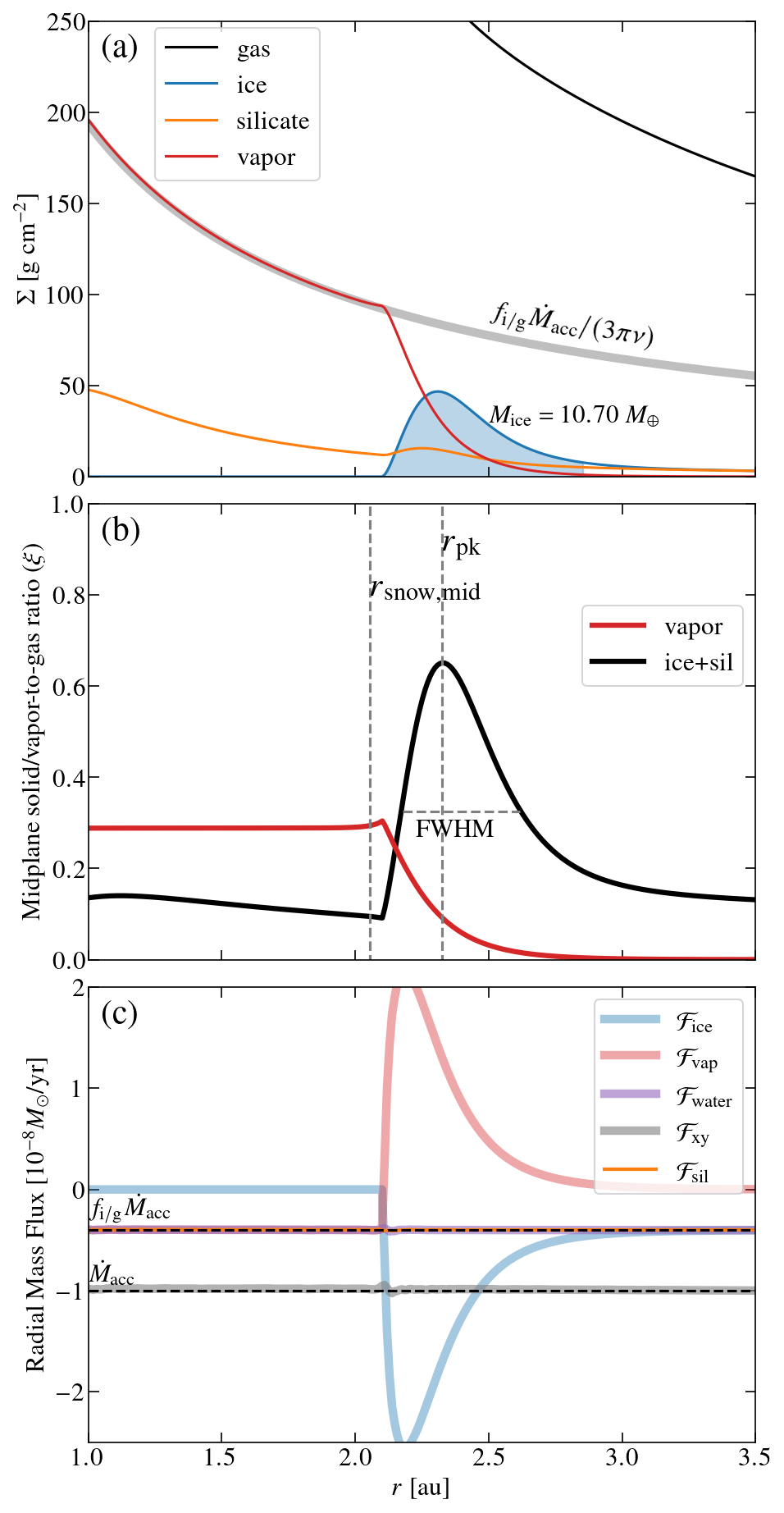}
	\caption{Radial profiles of the \texttt{passive} run. \textbf{(a):} Vertically-integrated surface density of total gas, ice, silicate and vapor. The grey line shows the expected vapor surface density from the unperturbed viscous solution (\eq{Sigma_g}), where $f_{\mathrm{i/g}}$ is the ice-to-gas flux ratio. The blue shade denotes \md{the region that contains} $M_{\mathrm{ice}}$. \textbf{(b):} The solid/vapor-to-gas ratio in the midplane. The positions of peak solid-to-gas ratio ($r_{\mathrm{pk}}$) and the midplane snowline ($r_{\mathrm{snow,mid}}$) are highlighted; \textbf{(c):} Vertically-integrated radial mass fluxes of ice, vapor, water, H-He and silicate, where $\mathcal{F}_{\mathrm{water}} = \mathcal{F}_{\mathrm{vap}} + \mathcal{F}_{\mathrm{ice}}$. The dashed lines denote the \md{imposed} gas ($\dot{M}_{\mathrm{acc}}$) and ice ($f_{\mathrm{i/g}}\dot{M}_{\mathrm{acc}}$) fluxes, respectively.
	}
	\label{fig:profile_iso_highL}
\end{figure}

\begin{figure*}
	\includegraphics[width=\textwidth]{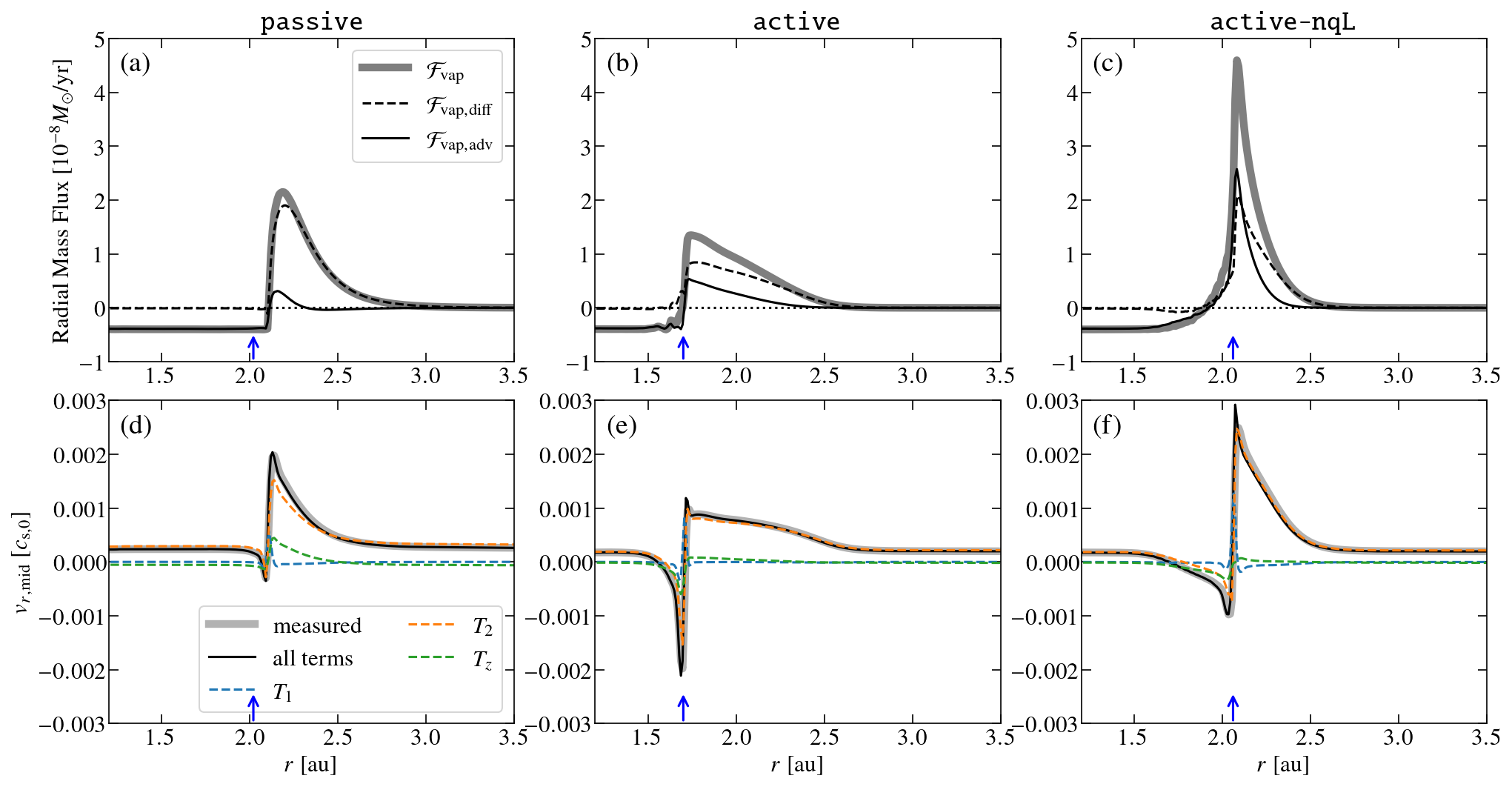}
	\caption{\textit{Upper panels:} Vertically-integrated radial mass flux of vapor. For three runs (\texttt{passive}, \texttt{active} and \texttt{active-nqL}), the diffusive, advective and total vapor flux are plotted. The dotted lines denote zero-flux levels. The blue arrows indicate the position of the midplane snowline ($r_{\mathrm{snow,mid}}$). The advective flux contributes significantly to the total vapor flux in the active disk runs (\texttt{active} and \texttt{active-nqL}), while it is negligible in the \texttt{passive} run. \textit{Lower panels:} Midplane gas radial velocity and the contributions from different terms following \eq{v_r}. The thick grey line represents the measured value from the simulation. The $T_2$ term signifies the dominant contribution of the radial viscous stress to the radial velocity, while the contributions from radial advection ($T_{1}$) and $z$-direction ($T_{z}$) are negligible.
    }
    \label{fig:v_r}
\end{figure*}

\subsection{Simulation setup and verification}
\label{sec:setup}
The midplane snowline is initially anchored at 2.1 au (\se{simu_cases}). We set up the simulation in a spherical-polar ($R$-$\theta$) coordinate system with the star centered at $R = 0$.  The simulation domain covers $R \in [1.0, 4.0]$ au and $\theta \in [1.3, \pi/2]$. We describe our models (\se{methods}) and conduct analysis of the data in a cylindrical coordinate ($r$-$z$) for convenience. At the reference position $r_{0} = 3.0~\mathrm{au}$ the simulation domain in $\theta$ corresponds to ${\approx}7 H_{\gas}$. The grid resolution is $N_{r} \times N_{\theta} = 320 \times 64$. To improve the resolution near the midplane region, where most of the material resides, we adopt a uniform spacing for $\theta^{1/3}$ (e.g., \citealt{OrmelEtal2015i}). This increases the grid resolution near the midplane by a factor of ${\approx}3$.

Simulations are initialized with non-condensible gas (H-He) and relaxed to a steady flow pattern. Then pebbles are released from the outer boundary and sublimate at the iceline region to enrich the inner disk with vapor. We monitor the ice and vapor fluxes ($\mathcal{F}_{\mathrm{ice}}$ and $\mathcal{F}_{\mathrm{vap}}$) and evolve the system until a steady state is reached. For all simulations, a steady state is reached after about $2 \times 10^{5}~\Omega^{-1}$. To illustrate the realization of the steady state, we plot the vertically-integrated radial mass flux of model \texttt{passive} in \fg{profile_iso_highL}c. Exterior to the iceline the ice flux $\mathcal{F}_\mathrm{ice}$ equals the imposed $f_{\mathrm{i/g}} \dot{M}_{\mathrm{acc}}$, until sublimation starts at around 2.5 au where the vapor flux becomes non-zero. In the snowline region, where phase change processes operate, the total water flux $\mathcal{F}_{\mathrm{water}} = \mathcal{F}_{\mathrm{ice}} + \mathcal{F}_{\mathrm{vap}}$ is conserved and equals the imposed ice flux. The silicate ($\mathcal{F}_{\mathrm{sil}}$) and H-He gas flux ($\mathcal{F}_{\mathrm{xy}}$) are also plotted in \fg{profile_iso_highL}c and shown to be conserved across the simulation domain.

\subsection{2D steady state structure and characteristic outputs of the solid pile-up}
\label{sec:2D_structure}
To illustrate the qualitative picture of the 2D snowline, we present the radial($r$)-vertical($z$) slices of the steady state density distribution in \fg{contourf_density} and the corresponding tempearture structure in \fg{contourf_Tem}. In these figures the green lines are drawn as the boundary $\rho_{\mathrm{p,ice}}/\rho_{\gas} = 10^{-2}$, which represent the surface where ice is exhausted in the disk (we call it ``snowline'' hereafter \footnote{Formally, snowline is defined following \eq{r_snow}. However, we use the surface $\rho_{\mathrm{p,ice}}/\rho_{\gas} = 10^{-2}$ because it closely follows the defined snowline at the ice sublimation region and highlights the scale height of pebbles.}). In \fg{contourf_density}, the region enclosed by the green line is filled with blue contours that represent the ice density \md{(normalized to the midplane gas density at $r_{0}$ of the unperturbed disk, i.e., $\rho_{0} \approx 3.4\times10^{-11}~\mathrm{cm~g^{-3}}$)}, while the rest is filled with red contours that represent the vapor density. To further visualize the water transport process, we plot ice (blue) and vapor (pink) mass fluxes density (in the unit of $\rho\bm{v}$) with streamlines, the thickness of which denote the strength of the mass flux. The mass flux density includes both the advective and diffusive components.

Consistent with previous 1D results (e.g., \citealt{SchoonenbergOrmel2017}), \fg{contourf_density} demonstrates the characteristic ice density bump exterior to the snowline in all runs.
This bump is more clearly illustrated in \fg{profile_iso_highL}a, where the surface density profile of the \texttt{passive} run is presented. The features observed in this run are representative to all simulations runs. \md{As pebbles drift in, the initially-equal surface density of ice and silicate ($\zeta = 0.5$) gradually separates due to the pile-up of ice. The pile-up in solids just exterior to the snowline is mostly contributed by ice rather than silicate (\fg{profile_iso_highL}a). Eventually all ice sublimates into vapor, which then follows the accretion flow of the gas. Therefore the surface density of vapor matches the expected viscous solution of gas in steady state (grey line), while silicate pebbles keep drifting towards the star at high velocity (typically 10 times larger than the gas accretion).} The corresponding midplane solid-to-gas ratio --- an indicator of planetesimal formation (e.g., \citealt{JohansenYoudin2007}) --- is plotted in \fg{profile_iso_highL}b. To quantify the properties of the solid pile-up and compare its strength among different simulation runs, we define the following indicators:
\begin{enumerate}
	\item The location of the midplane snowline $r_\mathrm{snow,mid}$ as defined in \eq{r_snow}. While $r_\mathrm{snow,mid}$ is initially anchored to 2.1 au, it could change after ice injection.
	\item The peak midplane solid-to-gas ratio $\xi_{\mathrm{pk}} = \max \left[(\rho_{\mathrm{p,ice}} + \rho_{\mathrm{p,sil}})/\rho_{\gas}\right]$
	      and its corresponding location $r_\mathrm{pk}$.
	\item The full-width-half-maximum (FWHM) of the solid pile-up as measured from the dust-to-gas ratio profile.
	\item The ice mass ($M_{\mathrm{ice}}$) stored in the solid pile-up. We integrate the ice surface density from the point where $\Sigma_{\mathrm{ice}}/\Sigma_{\mathrm{sil}} > 1.5$ inwards (blue-shaded region in \fg{profile_iso_highL}a).
\end{enumerate}
The positions of $r_{\mathrm{pk}}$, $r_{\mathrm{snow,mid}}$, and the FWHM of the solid pile-up are illustrated in \fg{profile_iso_highL}b and the value of these characteristic indicators are summarized in \tb{simu_cases} for all runs.

In previous 1D studies \citep{RosJohansen2013,DrazkowskaAlibert2017,SchoonenbergOrmel2017,HyodoEtal2021}, the solid pile-up is understood as a result of outward diffusion of vapor and recondensation.
To examine this picture, the vertically-integrated radial mass flux of \texttt{passive}, \texttt{active} and \texttt{active-nqL} are plotted in \fg{v_r} (upper panel). The vapor flux is further split into diffusive and advective components. Consistent with the flux density of vapor just exterior to the snowline shown in \fg{contourf_density}, we observe a strong outward flux of vapor in all three cases, responsible for the pile-up. In the \texttt{passive} run, the total outward vapor flux is dominated by diffusion. However, in the active disk runs (\texttt{active-nqL} and \texttt{active}), the contribution of the advective flux is comparable to that of the diffusive flux.
This implies that there must be a strong outflow near the snowline region, which is confirmed by the midplane radial gas velocity profile in \fg{v_r} (lower panel). While this radial outflow is automatically captured by the hydrodynamic simulation, 1D models relying on integrated versions of the transport equations (e.g., \citealt{SchoonenbergOrmel2017}) ignore it and therefore underestimate the pile-up strength (see the discussion in \app{1D_model}). In \se{gas_dynamics}, we present an analysis of this radial outflow to explain its origin.

Aside from the snowline region, water transport in general is characterized by the inward drift of ice from the outer boundary and the accretion of water vapor at the inner boundary (\fg{contourf_density}). Focusing on the disk region inside the snowline, where the mass flux is purely contributed by advection (see \fg{v_r}, upper panel), the streamlines of vapor flux density (pink) reveal outflow at the midplane and accretion at the upper regions (${\geq}1\,H_{\gas}$). \md{This characteristic flow pattern is consistent with previous studies of 2D viscous disk with pure H-He gas \citep{TakeuchiLin2002,Ciesla2009}.} The preferential gas accretion in the upper layer is due to the steep radial density gradient at the midplane, resulting in a net gain of angular momentum due to the greater viscous torque exerted by the faster-rotating inner disk compared to the slower-rotating outer disk (e.g., \citealt{TakeuchiLin2002}). Combined with the outflow in the midplane and the outward diffusion of vapor near the pile-up region, the simulations demonstrate that vapor leaves the snowline region and is accreted easier from higher altitude (see also \se{water_cycle}).

\subsection{Effect of snowline morphology}
\label{sec:morphology}
We now analyze the effect of the snowline morphology, first focusing on the simulation runs neglecting latent heat effects (\texttt{active-nqL}, \texttt{passive-nqL} and \texttt{iso-nqL}).
The temperature structures (\fg{contourf_Tem}) of the passive and isothermal disks looks similar, resulting in a similar shape of the snowline, which shifts inwards more at the midplane due to the higher ice density.
In contrast, the \texttt{active-nqL} run harbors a hotter midplane, resulting in an inward protrusion of the snowline above the midplane, similar to the findings by \citet{OkaEtal2011}.

Different snowline morphologies lead to distinct pile-up properties. From \tb{simu_cases}, the \texttt{active-nqL} run stores a significantly higher ice mass ($M_{\mathrm{ice}}$) in the pile-up region. The \texttt{active-nqL} run also has a narrower pile-up (FWHM), which results in a higher peak dust-to-gas ratio $\xi_\mathrm{pk}$. The smaller FWHM can be understood since the temperature gradient in the \texttt{active-nqL} run is larger across the snowline, as reflected by the closer-spaced contours in \fg{contourf_Tem}. As a result, the ice sublimation takes place in a narrower region.

Furthermore, from the ice flux density streamlines in \fg{contourf_density}a, we see that the \texttt{active-nqL} run is characterized by strong pebble settling in the pile-up region, a feature that is absent in the passive and isothermal disks. In all runs, ice sublimates mainly at the midplane near the snowline. The released vapor then spread upwards (mostly by diffusion). Since it is cooler in the upper layers in the active disk, vapor recondenses on pebbles, which then settle back to the midplane. This process is less efficient in the passive and isothermal disks since the temperature is more uniform in the vertical direction. As pebbles settle, they sublimate again, which leads to a ``cycle'' of pebble growth and settling. In steady state, we observe that the Stokes number of pebbles increases to ${\approx}0.1{-}0.2$ near the upper boundary of the snowline, contrasting the nominal value of ${\approx}0.03$ at the midplane, which leads to a strong downward flux of ice as seen in \fg{contourf_density}a.

As discussed in \se{2D_structure}, vapor escapes the snowline region predominantly from higher altitude. As a result, the recondensation of vapor in the upper layers of active disks effectively confines water to the snowline region. On the other hand, in the passive and isothermal disks, vapor can freely flow to the inner disk after it has spread to the upper layers. This water cycle observed in the active disk results in higher amounts of ice stored in the snowline region (\tb{simu_cases}). A further analysis will be presented in \se{water_cycle}.

\subsection{Effects on the rotational velocity}
\label{sec:headwind}
With vapor injection, the pressure support near the snowline will be altered due to the sudden increase of density and mean molecular weight, therefore changing the azimuthal velocity ($v_{\phi}$). The strength of the radial pressure gradient is usually expressed through \citep{NakagawaEtal1986},
\begin{equation}
	\label{eq:eta_define}
	\eta \equiv -\frac{1}{2} \frac{c_{\mathrm{s}}^{2}}{v_{\mathrm{k}}^{2}} \frac{\partial \log P_{\gas}}{ \partial \log r}
	\simeq \frac{v_{\mathrm{hw}}}{v_{\mathrm{k}}},
\end{equation}
where $v_{\mathrm{hw}} = v_{\mathrm{k}} - v_{\phi} \simeq \eta v_\mathrm{k}$ denotes the deviation of the gas motion from Keplerian motion, i.e., the headwind (e.g., \citealt{Armitage2020}). We measure $\eta$ at the midplane of the disk before vapor injection and after the steady state is reached (\fg{eta}).  Compared to disks without vapor injection (``unperturbed''), $\eta$ decreases for all runs inside the snowline. In addition, $\eta$ exhibits a dip right at the snowline location, $r_\mathrm{snow,mid}$.
To understand this feature, we expand the pressure gradient in \eq{eta_define} and write $\eta$ as,
\begin{equation}
	\label{eq:eta_decomp}
	\eta = -\frac{1}{2v_{\mathrm{k}}^{2}}
	\frac{k_{\mathrm{B}}T}{\mu m_{\mathrm{p}}} \frac{\mathrm{d}}{\mathrm{d} \log r} \left(\log \rho_{\gas} + \log T - \log \mu \right).
\end{equation}
The negative sign of the $\mu$-gradient term arises due to the decreasing pressure support from a higher mean molecular weight gas.

Inside the snowline, the decrease in the headwind is fully caused by the overall increase of $\mu$, i.e., the prefactor in \eq{eta_decomp}. Given that the water vapor fraction is $f_{\mathrm{v}} = f_{\mathrm{i/g}}/(1+f_{\mathrm{i/g}}) = 0.4/1.4$, $\mu$ will increase from $\mu_\mathrm{xy}=2.34$ to $\mu = 3.11$ (\eq{mmw}), consistent with the shifts of $\eta$ in \fg{eta}.
At the snowline, the dip in $\eta$ results from the steep \md{gradient} in the mean molecular weight ($-\mathrm{d} \log \mu / \mathrm{d} \log r$), which decreases pressure support, although the vapor injection ($\mathrm{d} \log \rho_{\gas} / \mathrm{d} \log r$) from ice sublimation counteracts this effect.
The dip is most pronounced in the \texttt{active-nqL} run where the vapor fraction increases steepest in the snowline region. Surprisingly, the \texttt{active} run exhibits a similar level of decrease in $\eta$ as the \texttt{active-nqL} run, despite its smaller gradient in $\mu$. The additional decrease in $\eta$ in the \texttt{active} run originates from the significant flattening of the temperature profile ($\log T$) by latent heat cooling, which will be discussed in \se{latent_heat}.

In summary, compared to pure H-He disk without vapor injection, the headwind generally decreases by a factor of $0.6-0.7$ at the snowline, which is significant enough to boost pebble accretion rates and promote SI (see \se{PA_onset}).

\begin{figure*}
	\includegraphics[width=2\columnwidth]{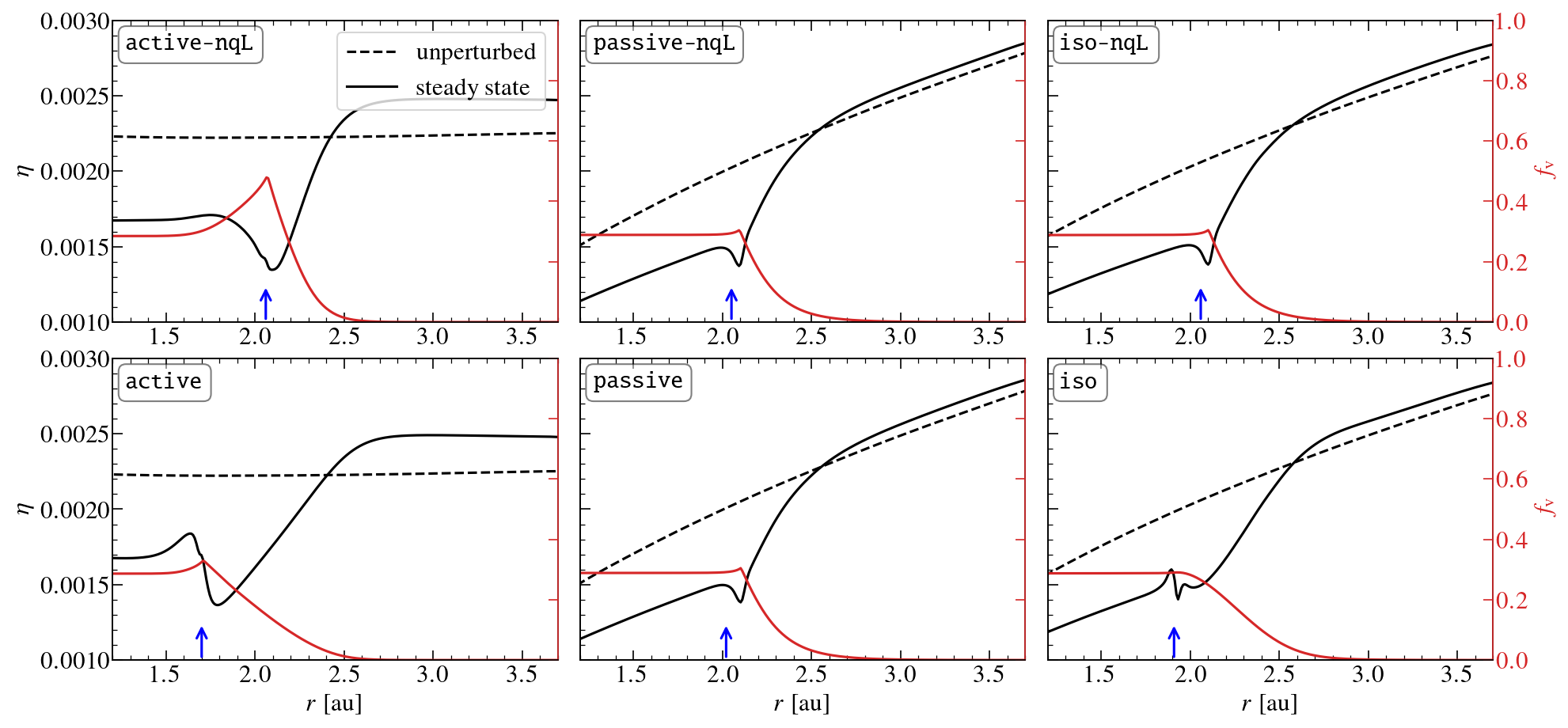}
	\caption{Azimuthal velocity deviation at the midplane\md{, expressed in terms of the radial pressure gradient parameter $\eta$ (\eq{eta_decomp})}. The black dashed and solid lines show $\eta$ before vapor injection and after the steady state is reached, respectively.
		The red lines show the vapor fraction corresponding to the y-axis on the right. The blue arrows indicate the position of the midplane snowline ($r_{\mathrm{snow,mid}}$).\md{A reduction of $\eta$ at the snowline region strengthens solid pileups and is conducive to planetesimal formation by streaming instability and planet formation by pebble accretion.}}
	\label{fig:eta}
\end{figure*}

\section{Effect of Latent heat exchange}
\label{sec:latent_heat}

\begin{figure}
	\includegraphics[width=\columnwidth]{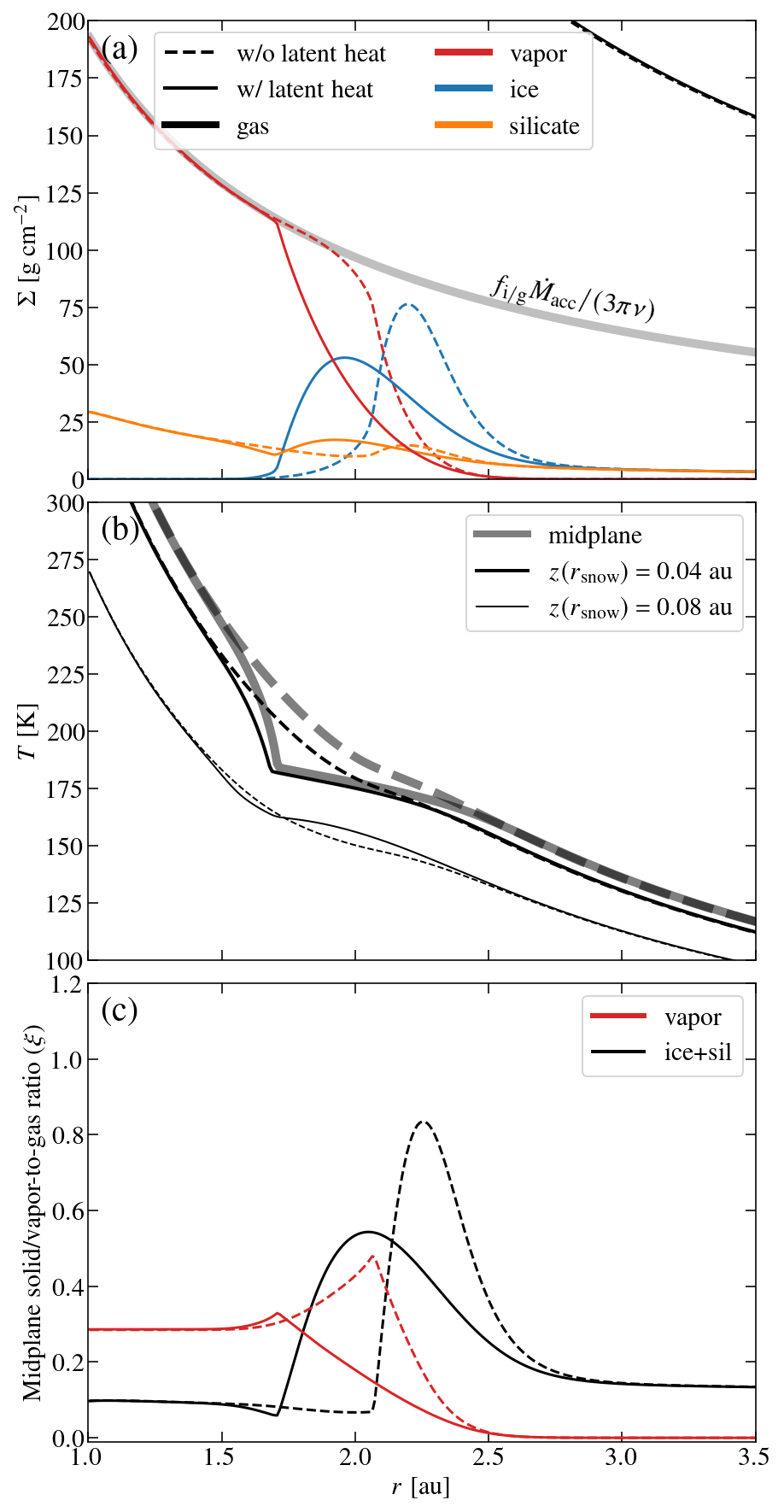}
	\caption{Comparisons of radial profiles of the active disks. For all three panels, the solid line represents the \texttt{active} disk while the dashed line represents \texttt{active-nqL} disk. \textbf{(a)}: The surface density profiles. As in \fg{profile_iso_highL}, the expected vapor surface density $f_{\mathrm{i/g}} \dot{M}_{\mathrm{acc}}/(3 \pi \nu)$ is plotted in grey. \textbf{(b)}: The temperature profile of slices in $\theta$-direction at the midplane and higher altitudes. The slices above the midplane pass through a certain height $z$ at the location of snowline ($r_{\mathrm{snow}}$) of the \texttt{active} run.  \textbf{(c)}: The solid-to-gas ratio (black) and vapor-to-gas ratio (red) at the midplane in steady state .}
	\label{fig:latent_contra}
\end{figure}

Latent heat exchange associated with ice sublimation and vapor recondensation has not been investigated explicitly in disks with hydro simulations before (see the discussion of \citealt{Owen2020}). To examine its impact on the thermal structure of the disk and the resulting solid pile-up, we run a series of simulations of different snowline morphology incorporating $q_{\mathrm{latent}}$, the latent heat exchange term. All other model parameters are kept the same (see \tb{simu_cases}) to enable a direct comparison.

\subsection{Active disks}
\label{sec:latent_active_disk}
In the active disk, the temperature structure near the snowline is significantly altered. Unlike the smooth, gradual temperature transition across the snowline seen in \texttt{active-nqL} (see the evenly-spaced contours in \fg{contourf_Tem}a), the \texttt{active} run (\fg{contourf_Tem}b) reveals a slowly-varying temperature ``plateau'' in the ice sublimation region, followed by a sharp temperature increase across the snowline (marked by green line). This is further illustrated in \fg{latent_contra}b, where a jump in the midplane temperature (solid line) is clearly noticeable.
The plateau arises from the latent heat absorption during ice sublimation. A maximum temperature difference of ${\approx}40$ K is observed when $q_{\mathrm{latent}}$ is included, highlighting the significant cooling effect due to latent heat absorption.

With the thermal structure altered by ice sublimation, the properties of the pile-up are also affected. First, the disk snowline is pushed inwards (see the green lines in \fg{contourf_Tem}a, b), resulting in a shift in the location of the pile-up (\fg{latent_contra}a). Second, both the width and the amplitude of the pile-up are modified. As shown in \fg{latent_contra}c, the temperature plateau created by latent heat cooling from ice sublimation broadens the snowline region. Consequently, the pile-up becomes around two-times more extended, while the peak solid-to-gas ratio is reduced by a factor of $\approx$1.6 (\tb{simu_cases}). The extended snowline region in the \texttt{active} run is also evident from \fg{contourf_density}b, where pebble settling happens over a wider region compared to the \texttt{active-nqL} run.

Comparing the temperature contours in \fg{contourf_Tem}a and b, we find that the latent heat exchange flattens the temperature gradient both in the vertical and radial directions. In \fg{latent_contra}b, we present the temperature profiles at various altitudes by selecting slices in $\theta$-direction.
For the midplane and $z(r_{\mathrm{snow}})=0.04~\mathrm{au}$ slices, the temperature is nearly uniform in the snowline region of the \texttt{active} run, whereas a distinct vertical temperature gradient is present in the \texttt{active-nqL} run. The more isothermal vertical temperature structure results from the combination of latent heat absorption due to ice sublimation at the hotter midplane and latent heat release from vapor recondensation at the cooler upper layers. As a result, ice settling is mitigated in the \texttt{active} run compared to the \texttt{active-nqL} run (\fg{contourf_density}, blue streamlines). However, accounting for latent heat exchange does not overturn the fundamental thermal stratification, i.e., the midplane remains hotter than the upper layers (\fg{contourf_Tem}b). As a result, the \texttt{active} disk exhibits a water cycle similar to that in the \texttt{active-nqL} disk (\fg{contourf_density}) and the total ice mass ($M_{\mathrm{ice}}$) stored in the snowline region is not significantly influenced by latent heat exchange (\tb{simu_cases}).

\subsection{Passive and isothermal disks}
\label{sec:latent_passive}
Although $q_{\mathrm{latent}}$ reshapes the snowline structures in active disks, its impact on passive and isothermal disks is significantly less pronounced, as evidenced by the characteristic indicators of the pile-up (\tb{simu_cases}). This is clearly seen by comparing the steady state thermal structure of runs with and without $q_{\mathrm{latent}}$ (\fg{contourf_Tem}). In active disks, the temperature contours are radially stretched and compressed over the range of around $1.7{-}2.3$ au at all altitudes within the phase change region. However, the temperature structure in \texttt{passive} is nearly identical to that in \texttt{passive-nqL}. In the isothermal disks, the temperature contours are only slightly bent near the snowline, implying a much weaker and more localized effect of latent heat.

These different effects of latent heat on the thermal structure can be explained by their distinct leading terms contributing to the temperature from \eq{T_z}. For passive disks in our simulations, due to the large $L_{\star}$ chosen (\tb{simu_cases}), the second (constant) term dominates, representing the radiation-equilibrium temperature determined by the stellar radiation field. Consequently, the contribution from $q_{\mathrm{latent}}$ is negligible.
Similarly, in isothermal disks, since it is very optically-thin ($\kappa_{\mathrm{R}}$ very low), the third term in \eq{T_z}, which represents the immediate balance between heating and cooling due to $q_{z}$, matters. This energy balance is local, therefore the latent heat effects are also localized.

In contrast, in active disks, where the snowline region is optically-thick (see the dotted lines in \fg{contourf_Tem}) and $L_{\star}$ is low, the "blanketing" effect, related to the effective, i.e., flux-weighted, optical depth $\tau_{\mathrm{eff}}$ (first term in \eq{T_z}), is dominant. Energy deposited deeper within the optically-thick disk heats more efficiently, as it takes longer for the energy to escape (e.g., \citealt{Chandrasekhar1935,MoriEtal2021}).
Since ice mainly sublimates at the midplane, latent heat absorption directly counteracts viscous heating, reducing the net heating ($q_z \ll q_\mathrm{vis}$), which significantly mitigates the ``blanketing'' effect.

To summarize, the effect of latent heat is most pronounced in disks where the blanketing effect is strong while it diminishes when cooling is efficient or when the disk is in radiation equilibrium.

\section{Analysis}
\label{sec:analysis}

\subsection{Gas dynamics near the snowline}
\label{sec:gas_dynamics}

As discussed in \se{2D_structure}, there is a strong outflow of gas near the snowline. This is clearly shown in \fg{v_r}, where the midplane gas velocity near the snowline is larger than other unperturbed region (e.g., ${\approx}10$ times larger in the \texttt{active-nqL} run). To understand it, we derive the expression for radial gas velocity in steady state, starting from the continuity and Navier-Stokes equations (considering angular momentum conservation) (e.g., \citealt{BalbusPapaloizou1999,Jacquet2013}). In cylindrical coordinate ($r$-$\phi$-$z$), the radial velocity of gas is given by,
\begin{equation}
	\label{eq:v_r}
	\begin{aligned}
		v_{r} = & -\frac{2}{r \rho v_{\mathrm{k}}} \frac{\partial}{\partial r}
		\left[
			\begin{matrix}
				\underbrace{r^{2}\rho v_{r} \delta v_{\phi}} \\ T_{1}
			\end{matrix}
			~
			\begin{matrix}
				\underbrace{+ \frac{3}{2} r \rho v_{\mathrm{k}} \nu - r^{3}\rho \nu \frac{\partial}{\partial r} \left(\frac{\delta v_{\phi}}{r} \right)} \\ T_{2}
			\end{matrix}
		\right]                                                                \\
		        &
		\begin{matrix}
			\underbrace{ -\frac{2}{\rho v_{\mathrm{k}}} \frac{\partial}{\partial z}\left[r \rho \delta v_{\phi} v_{z} - r \rho \nu \frac{\partial \delta v_{\phi}}{\partial z} \right]} \\ T_{z}
		\end{matrix}
	\end{aligned},
\end{equation}
where $v_{\mathrm{k}}$ represents the Keplerian velocity, $\delta v_{\phi}= v_{\phi} - v_{\mathrm{k}}$ is the deviation of azimuthal velocity from Keplerian and $v_{z}$ is the vertical velocity component. The steady-state radial velocity $v_{r}$ results from the angular momentum (AM) transport by advection and viscous stress in both radial and vertical directions.
The terms $T_{1}$ and $T_{2}$ represent the contributions from the radial advection and viscous stress, respectively, while $T_{z}$ encapsulates the AM transport from the $z$-direction as a whole.

Evaluating \eq{v_r}, \fg{v_r} shows the gas radial velocity $v_{r,\mathrm{mid}}$ in the midplane, where the outflow is the strongest. The contributions from each term are plotted and the sum of these terms matches the measured radial velocity, validating the decomposition. Across the snowline, there is a strong outward motion (positive) of gas followed by a sharp transition to inflow (negative). The term $T_{2}$ dominates over other terms in $v_{r,\mathrm{mid}}$, indicating that both outflow and inflow are primarily driven by viscous stress. This process can be understood as follows: as pebbles drift across the snowline and sublimate, the released vapor viscously spreads both outward and inward. In the absence of vapor injection, midplane gas motion is already characterized by outflow driven by the density gradient (see \se{2D_structure}). The injection of vapor into the system further steepens the initial radial density profile, exacerbating the outflow. \md{Compared with the background outflow inside the snowline, as identified in previous literature \citep{TakeuchiLin2002, Ciesla2009}, vapor injection induces {>}10 times stronger outflow at the snowline (\fg{v_r}, lower panel). } Furthermore, the strength of the viscous spreading is determined by the density of the deposited vapor. In the \texttt{active} disk, the latent heat cooling creates a temperature plateau (\fg{latent_contra}b), leading to a more gradual vapor deposition near the snowline (\fg{latent_contra}a), and consequently a weaker outflow compared to the \texttt{active-nqL} disk. Similarly, the inflow is stronger in the \texttt{active} disk because there is a steeper increase of temperature interior to $r_{\mathrm{snow,mid}}$, where ice is fully consumed and the latent heat cooling abruptly ceases. In summary, radial outflow from viscous spreading of vapor is commonly seen near the snowline.

\begin{figure*}
	\centering
	\includegraphics[width=2\columnwidth]{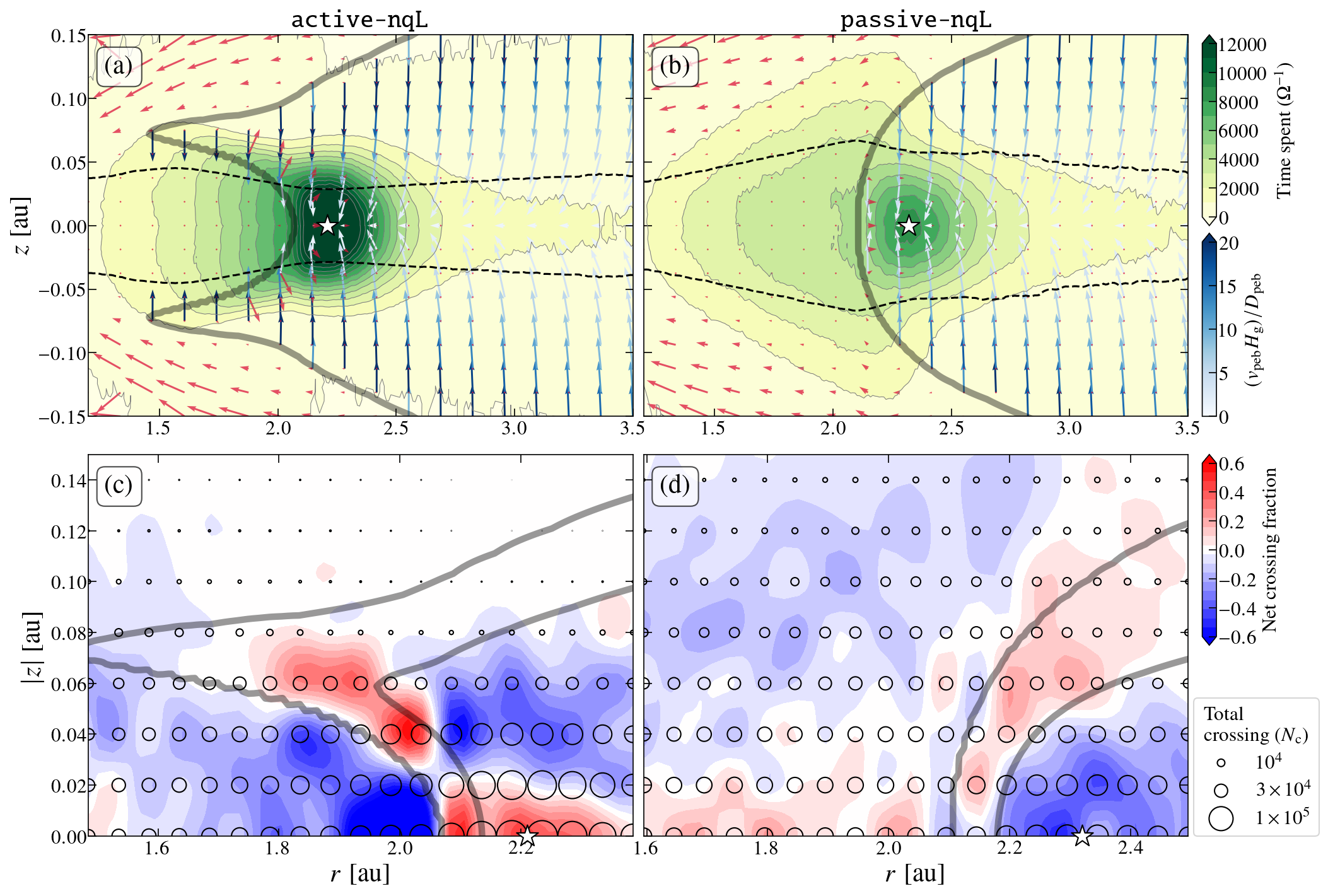}
	\caption{\textit{Upper panels}: Summary of the Lagrangian trajectory integration in the \texttt{active-nqL} and \texttt{passive-nqL} disks. Yellow-green color shading denotes the total time spent in each grid cell by the water particles, which are released from the outer boundary. The black dashed lines denote the root-mean-square of the positions of particles in the vertical direction. The stars denote $r_{\mathrm{pk}}$ of different runs (\tb{simu_cases}). The thick grey lines represents the location where $\rho_{\mathrm{p,ice}}/\rho_{\mathrm{vap}} = 1$ and $10^{-3}$, respectively, depicting the regions where particles are mostly in ice or vapor phase, respectively (see \app{lagrangian} for detailed explanation). The blue and red arrows represent the velocity of ice and vapor, respectively. The length of the arrows is proportional to $v H_{\gas} / D$ ($D, v$ are respectively the diffusivity and velocity of gas or pebble), which represent the ratio of the diffusion and advection timescales, $t_{\mathrm{dif}} / t_{\mathrm{adv}} \sim (H_{\gas}^{2} / D) / (H_{\gas}/v)$. We limit the length of the arrows so that so that $v H_{\gas} / D \leq 3$ for visualization purposes. Also, the depth of color of blue arrows represents $t_{\mathrm{dif,peb}} / t_{\mathrm{adv,peb}}$. \textit{Lower panels}: Crossing frequency of Lagrangian particles at various altitudes along the radial direction. Circles are plotted at the center of each vertical interval, with their surface area proportional to the total number of crossings ($N_{\mathrm{c}}$). Color indicates the ratio of net crossings to the total particle number ($N_{\mathrm{net}}/N_{\mathrm{p}}$). Here, negative (blue) values denote net accretion, whereas positive (red) values denote net outflow. The domain is zoomed-in to focus on the snowline region.
     \md{The checkerboard pattern in the net crossing fraction around the snowline seen in the \texttt{active-nqL} run is a manifestation of vigorous water cycling.}
	\label{fig:trace_contour}
    }
\end{figure*}

\subsection{Water-cycle enhances solid pile-up}
\label{sec:water_cycle}
In \se{morphology}, we identify a water-cycle that traps a large amount of ice in the snowline region in active disks. In other words, water parcels take longer time to escape the snowline region in active disks. To examine this finding, we perform Lagrangian trajectory integrations of water particles in the \texttt{active-nqL} and \texttt{passive-nqL} runs.
The trajectories of water particles---both in ice form as well as vapor--- is modelled with a Monte Carlo method, which includes a stochastic component, representing particle diffusion, as well as a systematic component, representing the steady-state advective velocity fields \citep{Ciesla2010,KrijtEtal2016}.
The simulation domain is mirrored to allow particles to travel into the lower hemisphere of the disk.
At any point, the phase of the water parcel is determined from the sublimation and condensation rates, which represent the probability to either transfer from ice to vapor or the other way around. See \app{lagrangian} for a detailed description.

For each disk, 1000 particles ($N_{\mathrm{p}}=1000$) are released from the outer boundary and integrated for $10^{5}~\Omega^{-1}$ or until the time where they get accreted ($r < 1.1$ au). The diffusion timescale for a water particle to spread over the gas scale height is $t_{\mathrm{dif}}\sim H_{\gas}^{2} / D_{\gas} \approx 300~\Omega^{-1}$. Therefore the integration time is sufficient for the particles to reach a diffusion-advection equilibrium state in the vertical direction. And as the radial transport time much exceeds the vertical transport time, trajectories are characterized by a large amount of vertical movement. After obtaining the trajectories, we count the time that particles spent in each grid cell ($t_{\mathrm{spend}}$) and plot its distribution in \fg{trace_contour}a,b. We observe a significant increase of $t_{\mathrm{spend}}$ near the snowline region, especially concentrated around the peak position of the pile-up ($r_{\mathrm{pk}}$). This consistency implies that the pile-up is indeed driven by longer escape time of particles.

Clearly, water particles spend significantly more time in the snowline region of the \texttt{active-nqL} disk compared to the \texttt{passive-nqL} disk, explaining the stronger pile-up of $M_{\mathrm{ice}}$ in active disks. \md{In other words, water is trapped in the active disks. Efficient transport of water away from the snowline region primarily occurs when water diffuses to upper layers ($z \approx 0.1$ au), where it is carried by the advective accretion flow (shown by the long red arrows in \fg{trace_contour}a,b). Near the midplane, however, only ``lucky'' particles can escape the snowline region by radial diffusion. This inefficiency arises because (1) vapor motion near the midplane ($z \approx 0-0.05$ au) is predominantly governed by diffusion, as depicted by the tiny red arrows in \fg{trace_contour}a,b, and (2) strong outflows along the snowline surface (thick grey lines), especially in the \texttt{active-nqL} run, create a significant barrier to the inward motion of water particles.}

\md{Moreover, due to the cold upper layer, strong ice settling (blue arrows in \fg{trace_contour}a,b), as part of the water-cycle, confines water much closer to the midplane in the \texttt{active-nqL} run. This is reflected in the root-mean-square of the particles' vertical positions ($\sqrt{\langle z^2\rangle}$) (indicated by black dashed lines in \fg{trace_contour}a,b), which represents the scaleheight of water parcels. In \fg{trace_contour}a, it is also evident that at the radial location where $t_\mathrm{spend}$ is particularly large at the midplane, the water scaleheight tends to be compressed. As a result, the key upper-layer channel to escape the snowline region is suppressed in the \texttt{active-nqL} run, leading to a stronger pile-up of ice.}
\md{The suppression of the upper-layer escape channel also explains why an outward, vertically integrated advective flux of vapor, comparable in magnitude to the diffusive flux, is observed in the active disk runs (\texttt{active} and \texttt{active-nqL}) but not in the \texttt{passive} run (\fg{v_r} and \se{2D_structure}). While all runs exhibit outflows that drive the outward advective flux of vapor at the midplane, the inward advective flux of vapor is absent in the upper layers of the active disk runs.}

To further illustrate this process, we examine the frequency with which particles cross specific radial slices near the snowline region.
We select several radial slices, spaced at 0.05 au intervals, that span both sides of $r_{\mathrm{pk}}$. Each slice is divided into multiple vertical intervals, and we record the number of particle crossings moving towards the star ($N_{-}$) and away from the star ($N_{+}$). The net crossing count at each slice, defined as $N_{\mathrm{net}} = N_{+} - N_{-}$, sums to the total number of particles, $N_{\mathrm{p}}$ (1000 in this case).

\begin{figure*}
	\sidecaption
	\includegraphics[width=0.7\textwidth]{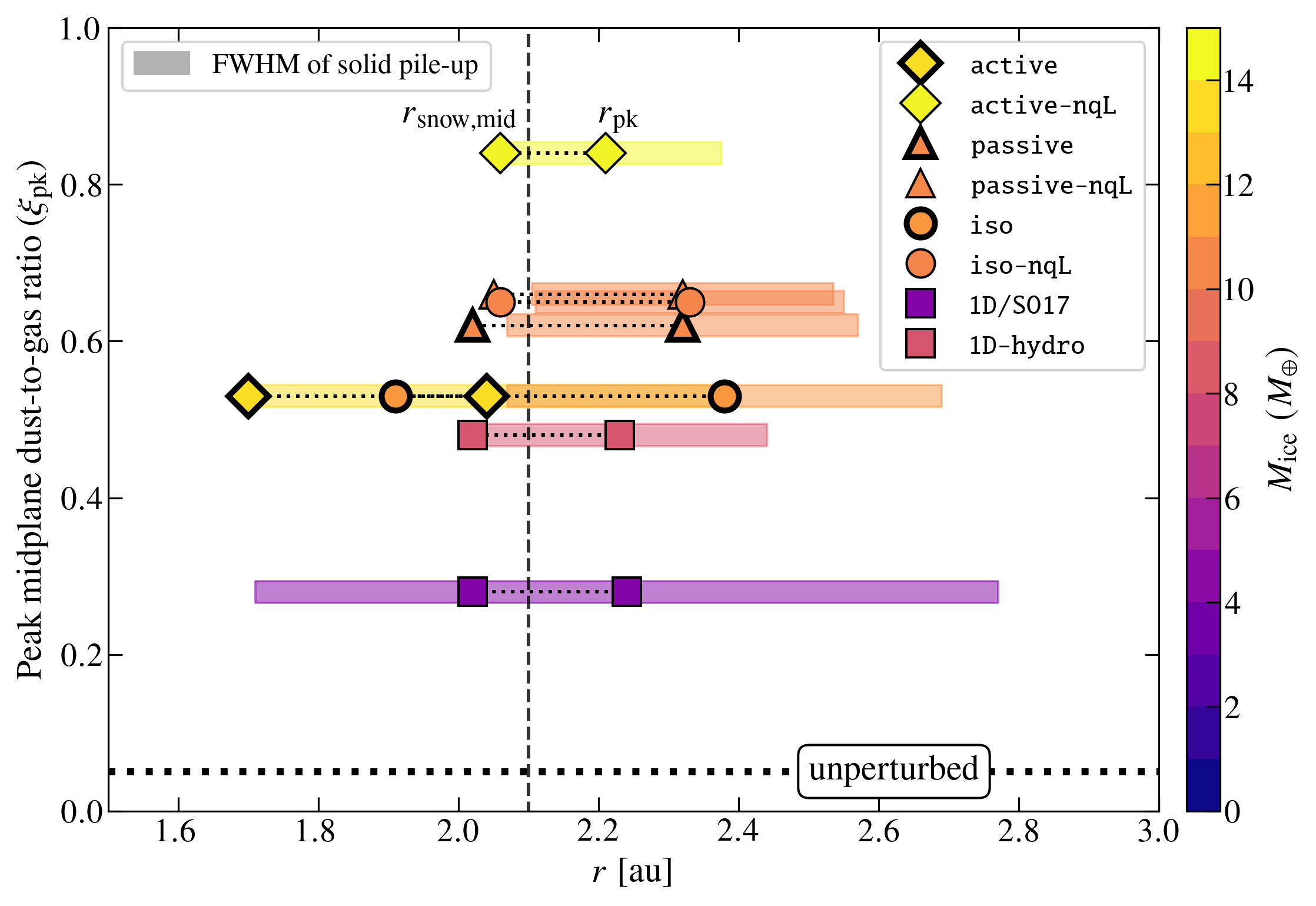}
	\caption{Summary of the pile-up properties of 1D and 2D runs as listed in \tb{simu_cases}. For each run, two markers are plotted to indicate the locations of the midplane snowline $r_{\mathrm{snow,mid}}$ (left) and the peak solid-to-gas ratio $r_{\mathrm{pk}}$ (right), with their $y$-axis indicating the value of the \md{peak} midplane solid-to-gas ratio $\xi_{\mathrm{pk}}$. \md{The vertical dashed line denotes the location of $r_{\mathrm{snow,mid}}$ before pebble injection (2.1 au). The horizontal dotted line, labeled as ``unperturbed'', denotes the expected solid-to-gas ratio at the snowline without ice sublimation.} The length of the shaded region, centered on $r_{\mathrm{pk}}$, denotes the FWHM of the pile-up. The color represents the total amount of water mass stored, as ice, in the snowline region.
    }
	\label{fig:output_summary}
\end{figure*}

In \fg{trace_contour}c and d, we focus on the snowline region of the \texttt{active-nqL} and \texttt{passive-nqL} runs. The net crossing fraction ($N_{\mathrm{net}}/N_{\mathrm{p}}$), shown in color, reveals two distinct mechanisms for water transport in the two scenarios. In the \texttt{active-nqL} case, water primarily exits the snowline through the midplane, via diffusion (indicated by the blue blob at 2.0 au), after crossing a barrier formed by the outflow (shown by the red band along the snowline). Conversely, in the \texttt{passive-nqL} run, water particles leave the snowline in addition through upper-layer accretion, as indicated by blue blobs at higher altitudes. The total number of crossing ($N_{\mathrm{c}}$), indicated by the size of the circles, peak around $r_{\mathrm{pk}}$ (the star symbols), with the \texttt{active-nqL} run exhibiting higher values than the \texttt{passive-nqL} run. This observation is consistent with \fg{trace_contour}a,b, indicating that the highly stochastic motion of water parcels near the snowline slows down accretion and that the ``upper-layer'' channel is significantly more efficient than the ``midplane'' channel.

\section{Discussion}
\label{sec:discussion}
In this section, we discuss the implications of the simulation results on planetesimal formation (\se{pltsml_formation}) and subsequent pebble accretion (\se{PA_onset}) in the snowline region. New observational features of snowline derived from the simulation is discussed in \se{obv_feature}.
Finally, we highlight the limitations of this study and propose potential directions for future research (\se{future_work}).

\subsection{Pile-up and planetesimal formation}
\label{sec:pltsml_formation}
Several studies have pointed out that a local solids-to-gas ratio of ${\sim}1$ is necessary to trigger SI. The SI would amplify the solids-to-gas ratio by orders of magnitude and result in planetesimal formation through gravitational collapse \citep{JohansenYoudin2007,BaiStone2010,CarreraEtal2015,LiYoudin2021}. The water snowline has long been suggested as a promising site to trigger SI due to its ability to enhance the midplane solids-to-gas ratios \citep{RosJohansen2013,DrazkowskaAlibert2017,SchoonenbergOrmel2017,HyodoEtal2019}.
\md{Our study specifically focuses on the ``vapor retro-diffusion'' scenario, which suggests the formation of icy planetesimals outside snowline, as the ``traffic jam''-induced pile-up is minor when pebble disaggregation is omitted (see \fg{profile_iso_highL}a and introduction). }

\Fg{output_summary} summarizes the outcome of the 2D simulations in terms of $\xi_\mathrm{pk}$. Two 1D runs are also included. In \texttt{1D/SO17} the gas velocity is fixed to the unperturbed viscous solution and in \texttt{1D-hydro} the gas motion is solved hydrodynamically (see \se{2D_structure} and \app{1D_model}). \md{For comparison, the estimated solid-to-gas ratio in an unperturbed disk \footnote{We estimate the solid-to-gas ratio by using $\xi \approx 
2f_{\mathrm{i/g}} v_{\gas} / v_{\mathrm{peb}}$ and evaluate it at $r_{0}$. The gas velocity follows viscous solution $v_{\gas} = 3\nu / (2r)$ and the drift velocity of pebble $v_{\mathrm{peb}} = (v_{\gas} + 2\eta v_{\mathrm{K}} \tau_{\mathrm{s}}) / (1+\tau_{\mathrm{s}}^{2})$ (e.g., \citealt{Weidenschilling1977,NakagawaEtal1986,Armitage2020}). }, where no ice sublimation takes place, is also highlighted. } \md{Already, in \texttt{1D/SO17}, the midplane solid-to-gas ratio is elevated at the snowline compared to the unperturbed disk due to the diffusion of vapor.} \md{From \fg{output_summary}} it can be seen that the resolved midplane outflow in the \texttt{1D-hydro} run \md{further} elevates $\xi_{\mathrm{pk}}$ by a factor of 1.7, highlighting the importance of the flow pattern.

Comparing the \texttt{1D-hydro} run with the 2D runs, the latter produce overall stronger pile-ups, as indicated by the larger $M_{\mathrm{ice}}$ and $\xi_{\mathrm{pk}}$. This highlights the importance of accounting for 2D flow patterns, where preferential accretion of vapor in the upper layers and settling of ice-rich pebbles establishes the water cycle across the snowline region, especially in active disks (\se{water_cycle}).
Among the 2D runs, the water ice mass stored in the pile-up region exhibits a clear difference (${\approx}1.5$ times) between the group of active disks and the group of isothermal and passive disks. This is due to a distinct snowline morphology and resulting distinct strength of the water cycle (see \se{water_cycle}). Though the inclusion of latent heat reduces $\xi_{\mathrm{pk}}$ and shifts $r_{\mathrm{snow,mid}}$ inwards, the water mass budget only changes slightly, implying that $M_{\mathrm{ice}}$ is mainly controlled by the snowline morphology.

None of the simulation runs reaches the values of $\xi_{\mathrm{pk}}$ as high as unity to trigger SI. However, inclusion of the backreaction of solids on the gas and accounting for disaggregation of ice-rich pebble after sublimation could boost the solid-to-gas ratio by factors ${\approx}4$ \citep{SchoonenbergOrmel2017} or even trigger runaway pile-up \citep{HyodoEtal2019}.  In addition, the headwind velocity is found to be reduced by a factor of ${\approx}0.6$ (\se{headwind}), which would bring down the metallicity threshold to trigger SI perhaps by a similar value \citep{BaiStone2010i,AbodEtal2019,BaronettEtal2024}.
However, this dependence is possibly non-linear and depends on pebble size \citep{BaiStone2010i}. Further study is needed to quantify the effect of headwind on SI to provide a complete picture of planetesimal formation at the snowline.

One key result of the hydrodynamical simulation in this work is the discovery of a water cycle, where water parcels are continuously exchanged between the solid and gas phases, enhancing the pile-up of solids (ice). The water-cycle is more pronounced in disks with a large vertical temperature gradient, i.e., active disks with efficient heating in the midplane and high opacity to preserve the heat.  Recently, young Class-0/I disks, have been found to have large dust scale height \citep{VillenaveEtal2023,LinEtal2023,Guerra-AlvaradoEtal2024} and systematically higher accretion rate (${\sim}10^{-7} M_{\odot}~\mathrm{yr}^{-1}$). They are therefore likely to host vigorous water cycles. \citet{HyodoEtal2021} proposes the preferential formation of giant planet's icy core during the early evolution stage when the disk snowline moves outward into the outer active zone (i.e., away from the inner MRI ``dead zone'', e.g., \citealt{Gammie1996}). They argue that the large diffusivity in the active zone facilitates outward diffusion of vapor and pile-up of ice exterior to the snowline. With the water cycle preferentially operated in young, active disks, our 2D results further support early formation of planetesimals.

\subsection{Onset of pebble accretion}
\label{sec:PA_onset}
After the formation of planetesimals, it has been suggested that  pebble accretion accelerates planetary growth \citep{OrmelKlahr2010,LambrechtsJohansen2012}.
Recently, the viability of pebble accretion in planetesimal rings has been explored in scenarios involving either ``clumpy rings'' where pebble drift slows down by the back-reaction on the gas \citep{JiangOrmel2023}, rings supported by pressure bumps \citep{Morbidelli2020i,GuileraEtal2020,LeeEtal2022}, or rings in smooth disks \citep{LiuEtal2019,JangEtal2022}.
To activate pebble accretion, planetesimals need to exceed a threshold mass \citep{OrmelKlahr2010,VisserOrmel2016,LiuJi2020}. Specifically, \citet{LiuOrmel2018} found that pebble accretion is fully activated at the mass $M_{*}=\eta^3 \tau_s m_\star$, which translates to a radius of,
\begin{equation}
	\begin{aligned}
		 & R_{*} = \left(\frac{\tau_{s} \eta^{3} M_{\star}}{4/3 \pi \rho_{\bullet,\mathrm{pltsml}}} \right)^{1/3}                                                                                                                                      \\
		 & \approx 308~\mathrm{km}~\left(\frac{\eta}{0.0013}\right) \left(\frac{\rho_{\bullet,\mathrm{pltsml}}}{1.0~\mathrm{g~cm^{-3}}}\right)^{-1/3} \left(\frac{\tau_{\mathrm{s}}}{0.03}\right)^{1/3}\left(\frac{M_{\star}}{M_{\odot}}\right)^{1/3}.
	\end{aligned}
\end{equation}
where
$\rho_{\bullet,\mathrm{pltsml}}$ is the internal density of planetesimal and $\eta = 0.0013$ corresponds to its minimum value at the snowline region in active disks (see \fg{eta}).
If this mass is not achieved by the planetsimal formation process, the planetesimals first need to grow via collisions in their birth belt \citep{LiuEtal2019}. However, N-body simulations find that the eccentricity and inclination of planetesimals are quickly excited by self-scattering, suppressing the accretion efficiency and making the collisional growth phase the main bottleneck to spawn planets (\citealt{LiuEtal2019,JangEtal2022,KaufmannEtal2025}).

Several studies have investigated the size distribution of planetesimals generated from SI \citep{SimonEtal2016,SchaeferEtal2017,AbodEtal2019,LiEtal2019}. Inspired by numerical simulations, \citet{LiuEtal2020} derives a general expression of the characteristic mass of planetesimals, which varies with the stellar mass and aspect ratio. Taking the solar mass and $h\approx0.04$ adopted from our simulation at $r_{\mathrm{pk}}$, the characteristic radius of planetesimals translates to ${\approx} 331~\mathrm{km} \gtrsim R_{*}$. Therefore, planetesimals formed at the snowline will immediately start efficient pebble accretion. In contrast, a value of $\eta \approx 0.0023$ (\fg{eta}) exterior to the snowline region, results in a much larger $R_{\ast} \approx 546~\mathrm{km}$ that falls short of the pebble accretion threshold. The reduction of $\eta$ due to vapor injection (\se{headwind}) therefore promotes planet formation at the snowline region.

Assuming the 3D limit, at $r_{\mathrm{pk}}$ these planetesimals would grow at a rate $\dot{m}_\mathrm{p} = 12 \tau_{\mathrm{s}} m_\mathrm{p} \rho_\mathrm{peb} \Omega_\mathrm{K} r^3 f_\mathrm{set}^2/M_\star$ \citep{OrmelLiu2018}, corresponding to a timescale,
\begin{equation}
	\label{eq:tpa3D}
	\begin{aligned}
		t_{\mathrm{PA,3D}} & = \frac{m_\mathrm{p}}{\dot{m}_\mathrm{p}} = 3.0 \times 10^{3} \left( \frac{r}{2.2\,\mathrm{au}} \right)^{-3/2} \left( \frac{\rho_\mathrm{peb}}{7\times10^{-11} ~ \mathrm{g~cm^{-3}}}\right)^{-1} \\
		                   & \times \left( \frac{\tau_{\mathrm{s}}}{0.03} \right)^{-1} \left( \frac{M_{\star}}{M_{\odot}} \right)^{1/2} \left( \frac{f_{\mathrm{set}}}{0.62} \right)^{-2}~\mathrm{yr}.
	\end{aligned}
\end{equation}
Here, $f_{\mathrm{set}}$ represents the settling fraction, with $f_{\mathrm{set}} \approx 0.62$ for a $331~\mathrm{km}$-sized planetesimal in our case. The pebble density, $\rho_{\mathrm{peb}} = \rho_{\mathrm{p,ice}} + \rho_{\mathrm{p,sil}}$, has its value measured from $r_{\mathrm{pk}}$ at the midplane of the \texttt{active-nqL} run.
This expression may break down at large $m_\mathrm{p}$ if accretion enters the 2D limit or when $\dot{m}_\mathrm{p}$ exceeds the pebble supply from the outer disk ($\dot{M}_\mathrm{peb}$). Nevertheless, \eq{tpa3D} shows that pebble accretion is rapid, leading to a mass doubling time of only several thousands of years. As a result, we expect the fast growth of a proto-Jupiter's icy core ($\approx 15 M_{\oplus}$) within $0.05~\mathrm{Myr}$ at the snowline region.

\subsection{Observational features of the water snowline}
\label{sec:obv_feature}
Despite the important role of the water snowline in planet formation, pinpointing its exact location within disks remains elusive (e.g., \citealt{Zhang2024}).
Around quiescent stars, it is extremely resolution-demanding to pinpoint the inner 3 au region \citep{Guerra-AlvaradoEtal2024i} (e.g., Taurus, $0.02''$ for 3 au at 150 pc) where the snowline is thought to reside.  Conversely, the outbursting FU Ori-type (FUor) systems, whose accretion rate can be elevated to ${\sim}10^{-4} M_{\odot}~\mathrm{yr}^{-1}$  \citep{FischerEtal2023}, will have extended their snowlines to 10--100 au, offering us a unique opportunity to hunt snowlines.
Intriguingly, V883 Ori, an FUor, shows an intensity depression (dark annulus) at ${\approx} 40 ~\mathrm{au}$ in ALMA band 6 continuum \citep{CiezaEtal2016}. Initially, this dip was interpreted as a change in dust density and opacity due to accumulation of small grains, which drift slower interior of the snowline.
This interpretation requires that the disk is optically thin at millimeter wavelength. On the other hand, \citet{KospalEtal2021} argued that FUors are massive, compact, and optically thick. In addition, the depression in line emission from rare isotopes (e.g., HDO, C$^{17}$O, \citealt{TobinEtal2023}) seen for the inner 40 au of V883 Ori, would imply optically-thick attenuation. Further insight comes from recent multi-wavelength analysis, which reveal that the spectral index remains continuous across the intensity depression, indicating that grain properties are not significantly altered across the dip \citep{HougeEtal2024}.
These findings suggest that the snowline region is optically thick and the intensity depression cannot be attributed to variations in dust properties.

These arguments leads to an appealing alternative explanation in line with the findings of this work: the observed intensity depression reflects a dip in temperature due to latent heat cooling. In \fg{latent_contra}b, it is observed that the temperature is reduced by ${\sim}20\%$ (40 K on top of 200 K) at the sublimation front, compared to model without latent heat effects, and that this temperature plateau affects a region of radial extent of about $50\% r_{\mathrm{snow,mid}}$. This reduction is sufficient to account for the intensity depression observed in V883 Ori (${\approx}15 \%$ in band 7, adopted from \citealt{HougeEtal2024}). For this scenario to hold, the sublimation front must extend vertically beyond the emission layer of the disk \md{(see the effect of emission layer in \citealt{van'tHoffEtal2018})}.

\md{While it is appealing to connect the intensity dip to latent heat, we caution that outbursting systems bear more nuances than quiescent systems. The accretion luminosity, and hence the disk temperature, evolves on a relatively short timescale ($\sim$century for FUors, \citealt{FischerEtal2023}), whereas the transport timescale is much longer at the outer region of the disks \citep{SchoonenbergEtal2017,ColmenaresEtal2024}. Therefore, it is likely that the outbursting systems are still evolving (non-steady) at the time of observation. The precise location of the moving snow surface further depends on the sublimation and condensation timescales (e.g., \citealt{SchoonenbergEtal2017}). From an observational perspective, there is also no consensus on the location of snowline in V883 Ori. Rather than being at 40 AU, water emission suggests it lies at $\approx 80$ au \citep{TobinEtal2023}. In future work, we would like to include these time-dependent aspects to understand the observational features of snowline in outbursting systems. }

\md{Additionally, the outburst itself could shape dust evolution and sublimation/condensation processes. After the outburst, the size distribution can deviate from collisional equilibrium for a long time \citep{HougeKrijt2023}. Depending on the cooling efficiency, condensation may occur preferentially on small silicate grains by heterogeneous nucleation rather than on ice-coated pebbles \citep{RosJohansen2024}. Future studies should assess the roles of the various physical processes during outburst to obtain a robust interpretation of current observations.  }

\subsection{Future directions}
\label{sec:future_work}
In this work, we have investigated the effect of snowline morphology and latent heat on snowline properties. Previous studies show that the mass accretion rate ($\dot{M}_{\mathrm{acc}}$), ice-to-gas flux ratio ($f_{\mathrm{i/g}}$) , disk viscosity (diffusivity), pebble \md{size} ($\tau_{\mathrm{s}}$) would all influence the pile-up strength, particularly the peak dust-to-gas ratio \citep{SchoonenbergOrmel2017,DrazkowskaAlibert2017,HyodoEtal2019}. How these parameters interact with key snowline processes identified in this work, such as the water cycle and latent heat cooling is non-trivial. \md{As the key parameter to drive flow pattern and active heating, we present results for different $\alpha$ in \app{vary_alpha}, demonstrating the robust existence of the water cycle. While an extensive parameter exploration has been omitted in this work, we discuss their potentially significant influence to guide future study.} For example, smaller-sized pebbles are found to enhance pile-ups due to their slower radial drift (e.g., \citealt{SchoonenbergOrmel2017}). With the larger scale height \md{and surface area} of smaller pebbles, the recapture of vapor could be more efficient. \md{However, the settling of pebbles will be less efficient, leading to uncertainty on the influence of the strength of water cycle when a realistic size distribution of pebbles is considered.} \md{Smaller pebbles are also more tightly coupled to the gas, resulting in a significantly reduced or even reversed drift velocity due to the outflow identified near the snowline. In this scenario, silicate grains can be a significant component of the pile-up by the ``outflow-induced'' traffic jam (see \app{vary_alpha}), in contrast to the ice-dominated pile-up found in this work (\se{2D_structure}). This outflow-induced traffic jam will be further exacerbated if pebble disintegration is considered (e.g., \citealt{HyodoEtal2019}). Additionally, coagulation is expected to occur efficiently in regions of solid pile-up (or dust rings), driven by high particle concentration and reduced drift speeds (e.g, \citealt{JiangEtal2024}).
	Understanding how the processes of coagulation and fragmentation shape the dust size distribution will be critical to determining water trapping efficiency near the snowline.}
Another key parameter is the ice-to-gas flux ratio $f_{\mathrm{i/g}}$.
Typically lower $f_{\mathrm{i/g}}$ leads to smaller $\xi_{\mathrm{pk}}$ of the solid pile-up (e.g., \citealt{SchoonenbergOrmel2017}). However, in the active disk, lowering $f_{\mathrm{i/g}}$ also mitigates the latent cooling effect that extends the width and reduces $\xi_{\mathrm{pk}}$ of the pile-up (\se{latent_active_disk}). Consequently, $\xi_{\mathrm{pk}}$ is not necessarily smaller with lower $f_{\mathrm{i/g}}$.
A thorough parameter exploration is necessary to arrive at a quantitative understanding and provide prescriptions for planet formation models relying on snowline.

The simulation setup in this work can be improved regarding its thermodynamic treatment. First, we assumed that the infrared opacity, which determines the cooling efficiency, is proportional to the non-condensible gas density (H and He) $\kappa = \kappa_{0} \rho_{\mathrm{H,He}}$ (\se{simu_cases}). In reality, icy mantle of dust grains could dominate the infrared opacity, given its higher abundance compared to silicates. Accounting for opacity from ice in the context of solar system's water snowline evolution, \citet{OkaEtal2011} find a slight outward shift of the snowline position due to an enhanced ``blanketing effect''.
Second, we assumed the standard $\alpha$-viscosity disk where the heating is deposited in the midplane.  However, in MHD simulations focusing on the inner disk region, \citet{MoriEtal2019} finds that the main heating source in the water snowline region is the Joule heating (due to the Ohmic resistivity of disk) that occurs at several gas scale heights above the midplane.  In contrast, the pebble scale height in dead zone could be ${\ll}H_{\gas}$ \citep{ZhuEtal2015,XuEtal2017,YangEtal2018}.  We therefore expect that the snowline morphology in MHD disks will resemble the isothermal/passive disks presented in this work, which should be confirmed in future simulations.
Since the influence of latent heat on temperature is weaker when heat is not deposited in the midplane (as discussed in \se{latent_passive}), we expect that the broadening of solid pile-up and the reduction of $\xi_{\mathrm{pk}}$ by latent heat are also weakened in MHD disks. Moreover, the flow structure in MHD disk is highly complex and different from viscous disk (e.g., \citealt{Bai2017}), which could drive different levels of pile-up, even becoming asymmetric about the midplane \citep{HuBai2021}.

While steady-state solutions always emerge in our simulations, it has recently been proposed that the CO snowline can become thermodynamically unstable when ice sublimation occurs \citep{Owen2020}. This instability arises when stellar heating is optically thick, but midplane cooling is optically thin, causing sublimation to reduce the cooling rate while the heating remains largely unaffected, leading to further temperature increases and more sublimation. The cooling conditions resemble the \texttt{iso} run in this work (\fg{contourf_Tem}, the midplane is optically-thin for cooling while $\kappa_{\mathrm{vi}} = 10~\mathrm{cm}^{-2}\mathrm{g}^{-1}$). However, since we do not account for opacity change due to ice sublimation, the instability cannot be triggered. In addition, latent heat exchange may operate against the instability given its cooling effect during ice sublimation. Although the latent heat effect is argued to be negligible in \citet{Owen2020}, it is observed to at least locally alter the temperature structure in the \texttt{iso} run (\fg{contourf_Tem}). Therefore, future 2D models incorporating opacity changes and latent heat effect should be conducted to confirm the feasibility of this snowline instability.

\section{Conclusions}
\label{sec:conclusion}
We conduct 2D multifluid hydrodynamic simulations towards the water snowline. The simulations include pebble dynamics, vapor transport, phase change processes while solving the temperature structures self-consistently with a two-stream radiative transfer method. We investigate how different snowline morphology and latent heat effects influence the solid pile-up outcome and explore the possible observational implications. Our main findings are:
\begin{enumerate}
	\item Viscous spreading of vapor released by inward-drifting ice-rich pebbles at the snowline drives a radial outflow, $\approx$10 times stronger than the background gas motion (\fg{v_r}). This previously unrecognized advective outward flux of vapor is of the same strength as the outward diffusive flux, augmenting the pile-up of ice outside the snowline.
	\item We identify a water cycle across the snowline region in active disks, where the temperature in the midplane is greater than the overlying layers (\fg{contourf_Tem}). Pebbles sublimate when passing through the snowline region near the midplane. The released vapor diffuses to cooler upper layers, where it recondenses on pebbles and contributes to their growth. These pebbles then settle and return the water back to the midplane (\se{morphology}).
	      This water cycle reduces the capability of water to escape from the snowline region via the upper layers, which, combined with the midplane outflow, significantly enhances the mass budget of ice in the snowline region (\se{water_cycle}). Therefore, active disks with a temperature inversion are more conducive to planetesimal formation than passive or isothermal disks.
	\item The total ice mass stored in the snowline region is mainly determined by the strength of the water cycle, which, in turn, depends on the morphology of the snowline (\fg{output_summary}). Radial temperature gradient across the snowline determines the extent and peak of ice pile-up, with a larger width and smaller peak dust-to-gas ratio under flatter temperature profiles.
	\item The release of vapor at the snowline reduces the azimuthal velocity of the gas in the disk ($\eta$ parameter; \fg{eta}), which lowers both the metallicity threshold for streaming instability and the threshold radius for planetesimals to accrete pebbles by a factor of ${\approx}0.6-0.7$. Planetesimals could immediately start pebble accretion after they are born in the snowline region. Due to the elevated densities, pebble accretion could occur rapidly with optimal growth timescale around 0.05 Myr for giant planets' icy core (\se{PA_onset}).
	\item The latent heat exchange tends to flatten the temperature gradient across the snowline, extending the width of the snowline region while reducing the peak solid-to-gas ratio (\fg{latent_contra}). These latent heat effects are most pronounced in active disks with strong blanketing effects.
	\item Latent heat cooling causes an extended plateau in the thermal structure of disks, up to $50\%$ of the snowline location ($r_{\mathrm{snow,mid}}$) in radial extent and ${\sim}40\,\mathrm{K}$ in temperature (\fg{latent_contra}). Such temperature dips would have observational imprints in the dust continuum emission at radio wavelengths (e.g., an intensity dip). In particular, outbursting systems, where the water snowline is pushed to ${\sim}10-100\,\mathrm{au}$, are promising targets to observationally test the predictions in this work with facilities such as ALMA.
\end{enumerate}

\begin{acknowledgements}
The authors thank the anonymous referee for providing a thoughtful and helpful report.
	YW acknowledges helpful discussions with Enrique Macias, Haochang Jiang, Wenrui Xu, Ruobing Dong, Rolf Kuiper, Mario Flock and Anders Johansen.
    This work is supported by the National Natural Science Foundation of China under grant Nos. 12250610189, 12233004 and 12473065.
	SM acknowledges support by Japan Society for the Promotion of Science KAKENHI grant No. 22K14081.
    The authors acknowledge the Tsinghua Astrophysics High-Performance Computing
platform at Tsinghua University for providing computational and data storage resources that have contributed to the research results reported within this paper.
\end{acknowledgements}

%
%

\bibliographystyle{aa} 
\bibliography{ads} 

\begin{thebibliography}{113}
\expandafter\ifx\csname natexlab\endcsname\relax\def\natexlab#1{#1}\fi

\bibitem[{{Abod} {et~al.}(2019){Abod}, {Simon}, {Li}, {Armitage}, {Youdin}, \& {Kretke}}]{AbodEtal2019}
{Abod}, C.~P., {Simon}, J.~B., {Li}, R., {et~al.} 2019, \apj, 883, 192

\bibitem[{{Armitage}(2020)}]{Armitage2020}
{Armitage}, P.~J. 2020, {Astrophysics of planet formation, Second Edition}

\bibitem[{{Aumatell} \& {Wurm}(2011)}]{AumatellWurm2011}
{Aumatell}, G. \& {Wurm}, G. 2011, \mnras, 418, L1

\bibitem[{{Bai}(2017)}]{Bai2017}
{Bai}, X.-N. 2017, \apj, 845, 75

\bibitem[{{Bai} \& {Stone}(2010{\natexlab{a}})}]{BaiStone2010}
{Bai}, X.-N. \& {Stone}, J.~M. 2010{\natexlab{a}}, \apj, 722, 1437

\bibitem[{{Bai} \& {Stone}(2010{\natexlab{b}})}]{BaiStone2010i}
{Bai}, X.-N. \& {Stone}, J.~M. 2010{\natexlab{b}}, \apjl, 722, L220

\bibitem[{{Balbus} \& {Papaloizou}(1999)}]{BalbusPapaloizou1999}
{Balbus}, S.~A. \& {Papaloizou}, J. C.~B. 1999, \apj, 521, 650

\bibitem[{{Baronett} {et~al.}(2024){Baronett}, {Yang}, \& {Zhu}}]{BaronettEtal2024}
{Baronett}, S.~A., {Yang}, C.-C., \& {Zhu}, Z. 2024, \mnras, 529, 275

\bibitem[{{Birnstiel}(2024)}]{Birnstiel2024}
{Birnstiel}, T. 2024, \araa, 62, 157

\bibitem[{{Birnstiel} {et~al.}(2010){Birnstiel}, {Dullemond}, \& {Brauer}}]{BirnstielEtal2010}
{Birnstiel}, T., {Dullemond}, C.~P., \& {Brauer}, F. 2010, \aap, 513, A79

\bibitem[{{Birnstiel} {et~al.}(2018){Birnstiel}, {Dullemond}, {Zhu}, {Andrews}, {Bai}, {Wilner}, {Carpenter}, {Huang}, {Isella}, {Benisty}, {P{\'e}rez}, \& {Zhang}}]{BirnstielEtal2018}
{Birnstiel}, T., {Dullemond}, C.~P., {Zhu}, Z., {et~al.} 2018, \apjl, 869, L45

\bibitem[{{Birnstiel} {et~al.}(2012){Birnstiel}, {Klahr}, \& {Ercolano}}]{BirnstielEtal2012}
{Birnstiel}, T., {Klahr}, H., \& {Ercolano}, B. 2012, \aap, 539, A148

\bibitem[{{Blum} \& {Wurm}(2000)}]{BlumWurm2000}
{Blum}, J. \& {Wurm}, G. 2000, \icarus, 143, 138

\bibitem[{{Blum} \& {Wurm}(2008)}]{BlumWurm2008}
{Blum}, J. \& {Wurm}, G. 2008, \araa, 46, 21

\bibitem[{{Calvet} {et~al.}(1991){Calvet}, {Patino}, {Magris}, \& {D'Alessio}}]{CalvetEtal1991}
{Calvet}, N., {Patino}, A., {Magris}, G.~C., \& {D'Alessio}, P. 1991, \apj, 380, 617

\bibitem[{{Carrera} {et~al.}(2015){Carrera}, {Johansen}, \& {Davies}}]{CarreraEtal2015}
{Carrera}, D., {Johansen}, A., \& {Davies}, M.~B. 2015, \aap, 579, A43

\bibitem[{{Chandrasekhar}(1935)}]{Chandrasekhar1935}
{Chandrasekhar}, S. 1935, \mnras, 96, 21

\bibitem[{{Chapman} \& {Cowling}(1991)}]{ChapmanCowling1991}
{Chapman}, S. \& {Cowling}, T.~G. 1991, {The Mathematical Theory of Non-uniform Gases}

\bibitem[{{Chiang} \& {Goldreich}(1997)}]{ChiangGoldreich1997}
{Chiang}, E.~I. \& {Goldreich}, P. 1997, \apj, 490, 368

\bibitem[{{Ciesla}(2009)}]{Ciesla2009}
{Ciesla}, F.~J. 2009, \icarus, 200, 655

\bibitem[{{Ciesla}(2010)}]{Ciesla2010}
{Ciesla}, F.~J. 2010, \apj, 723, 514

\bibitem[{{Cieza} {et~al.}(2016){Cieza}, {Casassus}, {Tobin}, {Bos}, {Williams}, {Perez}, {Zhu}, {Caceres}, {Canovas}, {Dunham}, {Hales}, {Prieto}, {Principe}, {Schreiber}, {Ruiz-Rodriguez}, \& {Zurlo}}]{CiezaEtal2016}
{Cieza}, L.~A., {Casassus}, S., {Tobin}, J., {et~al.} 2016, \nat, 535, 258

\bibitem[{{Colmenares} {et~al.}(2024){Colmenares}, {Lambrechts}, {van Kooten}, \& {Johansen}}]{ColmenaresEtal2024}
{Colmenares}, M.~J., {Lambrechts}, M., {van Kooten}, E., \& {Johansen}, A. 2024, \aap, 685, A114

\bibitem[{{Cuzzi} \& {Zahnle}(2004)}]{CuzziZahnle2004}
{Cuzzi}, J.~N. \& {Zahnle}, K.~J. 2004, \apj, 614, 490

\bibitem[{{Dominik} \& {Tielens}(1997)}]{DominikTielens1997}
{Dominik}, C. \& {Tielens}, A.~G.~G.~M. 1997, \apj, 480, 647

\bibitem[{{Dr{\k{a}}{\.z}kowska} \& {Alibert}(2017)}]{DrazkowskaAlibert2017}
{Dr{\k{a}}{\.z}kowska}, J. \& {Alibert}, Y. 2017, \aap, 608, A92

\bibitem[{{Dr{\k{a}}{\.z}kowska} {et~al.}(2023){Dr{\k{a}}{\.z}kowska}, {Bitsch}, {Lambrechts}, {Mulders}, {Harsono}, {Vazan}, {Liu}, {Ormel}, {Kretke}, \& {Morbidelli}}]{DrazkowskaEtal2023}
{Dr{\k{a}}{\.z}kowska}, J., {Bitsch}, B., {Lambrechts}, M., {et~al.} 2023, in Astronomical Society of the Pacific Conference Series, Vol. 534, Protostars and Planets VII, ed. S.~{Inutsuka}, Y.~{Aikawa}, T.~{Muto}, K.~{Tomida}, \& M.~{Tamura}, 717

\bibitem[{{Fischer} {et~al.}(2023){Fischer}, {Hillenbrand}, {Herczeg}, {Johnstone}, {Kospal}, \& {Dunham}}]{FischerEtal2023}
{Fischer}, W.~J., {Hillenbrand}, L.~A., {Herczeg}, G.~J., {et~al.} 2023, in Astronomical Society of the Pacific Conference Series, Vol. 534, Protostars and Planets VII, ed. S.~{Inutsuka}, Y.~{Aikawa}, T.~{Muto}, K.~{Tomida}, \& M.~{Tamura}, 355

\bibitem[{{Fray} \& {Schmitt}(2009)}]{FraySchmitt2009}
{Fray}, N. \& {Schmitt}, B. 2009, \planss, 57, 2053

\bibitem[{{Gammie}(1996)}]{Gammie1996}
{Gammie}, C.~F. 1996, \apj, 457, 355

\bibitem[{{Gillon} {et~al.}(2016){Gillon}, {Jehin}, {Lederer}, {Delrez}, {de Wit}, {Burdanov}, {Van Grootel}, {Burgasser}, {Triaud}, {Opitom}, {Demory}, {Sahu}, {Bardalez Gagliuffi}, {Magain}, \& {Queloz}}]{GillonEtal2016}
{Gillon}, M., {Jehin}, E., {Lederer}, S.~M., {et~al.} 2016, \nat, 533, 221

\bibitem[{{Gillon} {et~al.}(2017){Gillon}, {Triaud}, {Demory}, {Jehin}, {Agol}, {Deck}, {Lederer}, {de Wit}, {Burdanov}, {Ingalls}, {Bolmont}, {Leconte}, {Raymond}, {Selsis}, {Turbet}, {Barkaoui}, {Burgasser}, {Burleigh}, {Carey}, {Chaushev}, {Copperwheat}, {Delrez}, {Fernandes}, {Holdsworth}, {Kotze}, {Van Grootel}, {Almleaky}, {Benkhaldoun}, {Magain}, \& {Queloz}}]{GillonEtal2017}
{Gillon}, M., {Triaud}, A. H.~M.~J., {Demory}, B.-O., {et~al.} 2017, \nat, 542, 456

\bibitem[{{Guerra-Alvarado} {et~al.}(2024{\natexlab{a}}){Guerra-Alvarado}, {Carrasco-Gonz{\'a}lez}, {Mac{\'\i}as}, {van der Marel}, {Houge}, {Maud}, {Pinilla}, {Villenave}, {Asaki}, \& {Humphreys}}]{Guerra-AlvaradoEtal2024i}
{Guerra-Alvarado}, O.~M., {Carrasco-Gonz{\'a}lez}, C., {Mac{\'\i}as}, E., {et~al.} 2024{\natexlab{a}}, \aap, 686, A298

\bibitem[{{Guerra-Alvarado} {et~al.}(2024{\natexlab{b}}){Guerra-Alvarado}, {van der Marel}, {Di Francesco}, {Looney}, {Tobin}, {Cox}, {Sheehan}, {Wilner}, {Mac{\'\i}as}, \& {Carrasco-Gonz{\'a}lez}}]{Guerra-AlvaradoEtal2024}
{Guerra-Alvarado}, O.~M., {van der Marel}, N., {Di Francesco}, J., {et~al.} 2024{\natexlab{b}}, \aap, 681, A82

\bibitem[{{Guilera} {et~al.}(2020){Guilera}, {S{\'a}ndor}, {Ronco}, {Venturini}, \& {Miller Bertolami}}]{GuileraEtal2020}
{Guilera}, O.~M., {S{\'a}ndor}, Z., {Ronco}, M.~P., {Venturini}, J., \& {Miller Bertolami}, M.~M. 2020, \aap, 642, A140

\bibitem[{{Gundlach} \& {Blum}(2015)}]{GundlachBlum2015}
{Gundlach}, B. \& {Blum}, J. 2015, \apj, 798, 34

\bibitem[{{Gundlach} {et~al.}(2018){Gundlach}, {Schmidt}, {Kreuzig}, {Bischoff}, {Rezaei}, {Kothe}, {Blum}, {Grzesik}, \& {Stoll}}]{GundlachEtal2018}
{Gundlach}, B., {Schmidt}, K.~P., {Kreuzig}, C., {et~al.} 2018, \mnras, 479, 1273

\bibitem[{{Houge} \& {Krijt}(2023)}]{HougeKrijt2023}
{Houge}, A. \& {Krijt}, S. 2023, \mnras, 521, 5826

\bibitem[{{Houge} {et~al.}(2024){Houge}, {Mac{\'\i}as}, \& {Krijt}}]{HougeEtal2024}
{Houge}, A., {Mac{\'\i}as}, E., \& {Krijt}, S. 2024, \mnras, 527, 9668

\bibitem[{{Hu} \& {Bai}(2021)}]{HuBai2021}
{Hu}, Z. \& {Bai}, X.-N. 2021, \mnras, 503, 162

\bibitem[{{Huang} {et~al.}(2018){Huang}, {Andrews}, {P{\'e}rez}, {Zhu}, {Dullemond}, {Isella}, {Benisty}, {Bai}, {Birnstiel}, {Carpenter}, {Guzm{\'a}n}, {Hughes}, {{\"O}berg}, {Ricci}, {Wilner}, \& {Zhang}}]{HuangEtal2018}
{Huang}, J., {Andrews}, S.~M., {P{\'e}rez}, L.~M., {et~al.} 2018, \apjl, 869, L43

\bibitem[{{Huang} \& {Bai}(2022)}]{HuangBai2022}
{Huang}, P. \& {Bai}, X.-N. 2022, \apjs, 262, 11

\bibitem[{{Hubeny}(1990)}]{Hubeny1990}
{Hubeny}, I. 1990, \apj, 351, 632

\bibitem[{{Hyodo} {et~al.}(2021){Hyodo}, {Guillot}, {Ida}, {Okuzumi}, \& {Youdin}}]{HyodoEtal2021}
{Hyodo}, R., {Guillot}, T., {Ida}, S., {Okuzumi}, S., \& {Youdin}, A.~N. 2021, \aap, 646, A14

\bibitem[{{Hyodo} {et~al.}(2019){Hyodo}, {Ida}, \& {Charnoz}}]{HyodoEtal2019}
{Hyodo}, R., {Ida}, S., \& {Charnoz}, S. 2019, \aap, 629, A90

\bibitem[{{Ida} {et~al.}(2016){Ida}, {Guillot}, \& {Morbidelli}}]{IdaEtal2016}
{Ida}, S., {Guillot}, T., \& {Morbidelli}, A. 2016, \aap, 591, A72

\bibitem[{{Jacquet}(2013)}]{Jacquet2013}
{Jacquet}, E. 2013, \aap, 551, A75

\bibitem[{{Jang} {et~al.}(2022){Jang}, {Liu}, \& {Johansen}}]{JangEtal2022}
{Jang}, H., {Liu}, B., \& {Johansen}, A. 2022, \aap, 664, A86

\bibitem[{{Jiang} {et~al.}(2024){Jiang}, {Mac{\'\i}as}, {Guerra-Alvarado}, \& {Carrasco-Gonz{\'a}lez}}]{JiangEtal2024}
{Jiang}, H., {Mac{\'\i}as}, E., {Guerra-Alvarado}, O.~M., \& {Carrasco-Gonz{\'a}lez}, C. 2024, \aap, 682, A32

\bibitem[{{Jiang} \& {Ormel}(2023)}]{JiangOrmel2023}
{Jiang}, H. \& {Ormel}, C.~W. 2023, \mnras, 518, 3877

\bibitem[{{Jiang} {et~al.}(2023){Jiang}, {Wang}, {Ormel}, {Krijt}, \& {Dong}}]{JiangEtal2023}
{Jiang}, H., {Wang}, Y., {Ormel}, C.~W., {Krijt}, S., \& {Dong}, R. 2023, \aap, 678, A33

\bibitem[{{Johansen} {et~al.}(2007){Johansen}, {Oishi}, {Mac Low}, {Klahr}, {Henning}, \& {Youdin}}]{JohansenEtal2007}
{Johansen}, A., {Oishi}, J.~S., {Mac Low}, M.-M., {et~al.} 2007, \nat, 448, 1022

\bibitem[{{Johansen} {et~al.}(2021){Johansen}, {Ronnet}, {Bizzarro}, {Schiller}, {Lambrechts}, {Nordlund}, \& {Lammer}}]{JohansenEtal2021}
{Johansen}, A., {Ronnet}, T., {Bizzarro}, M., {et~al.} 2021, Science Advances, 7, eabc0444

\bibitem[{{Johansen} \& {Youdin}(2007)}]{JohansenYoudin2007}
{Johansen}, A. \& {Youdin}, A. 2007, \apj, 662, 627

\bibitem[{{Johansen} {et~al.}(2009){Johansen}, {Youdin}, \& {Mac Low}}]{JohansenEtal2009}
{Johansen}, A., {Youdin}, A., \& {Mac Low}, M.-M. 2009, \apjl, 704, L75

\bibitem[{{Kaufmann} {et~al.}(2025){Kaufmann}, {Guilera}, {Alibert}, \& {San Sebasti{\'a}n}}]{KaufmannEtal2025}
{Kaufmann}, N., {Guilera}, O.~M., {Alibert}, Y., \& {San Sebasti{\'a}n}, I.~L. 2025, arXiv e-prints, arXiv:2502.02124

\bibitem[{{Kippenhahn} {et~al.}(2013){Kippenhahn}, {Weigert}, \& {Weiss}}]{KippenhahnEtal2013}
{Kippenhahn}, R., {Weigert}, A., \& {Weiss}, A. 2013, {Stellar Structure and Evolution}

\bibitem[{{K{\'o}sp{\'a}l} {et~al.}(2021){K{\'o}sp{\'a}l}, {Cruz-S{\'a}enz de Miera}, {White}, {{\'A}brah{\'a}m}, {Chen}, {Csengeri}, {Dong}, {Dunham}, {Feh{\'e}r}, {Green}, {Hashimoto}, {Henning}, {Hogerheijde}, {Kudo}, {Liu}, {Takami}, \& {Vorobyov}}]{KospalEtal2021}
{K{\'o}sp{\'a}l}, {\'A}., {Cruz-S{\'a}enz de Miera}, F., {White}, J.~A., {et~al.} 2021, \apjs, 256, 30

\bibitem[{{Krijt} {et~al.}(2016){Krijt}, {Ormel}, {Dominik}, \& {Tielens}}]{KrijtEtal2016}
{Krijt}, S., {Ormel}, C.~W., {Dominik}, C., \& {Tielens}, A.~G.~G.~M. 2016, \aap, 586, A20

\bibitem[{{Lambrechts} \& {Johansen}(2012)}]{LambrechtsJohansen2012}
{Lambrechts}, M. \& {Johansen}, A. 2012, \aap, 544, A32

\bibitem[{{Lambrechts} \& {Johansen}(2014)}]{LambrechtsJohansen2014}
{Lambrechts}, M. \& {Johansen}, A. 2014, \aap, 572, A107

\bibitem[{{Lee} {et~al.}(2022){Lee}, {Fuentes}, \& {Hopkins}}]{LeeEtal2022}
{Lee}, E.~J., {Fuentes}, J.~R., \& {Hopkins}, P.~F. 2022, \apj, 937, 95

\bibitem[{{Li} \& {Youdin}(2021)}]{LiYoudin2021}
{Li}, R. \& {Youdin}, A.~N. 2021, \apj, 919, 107

\bibitem[{{Li} {et~al.}(2019){Li}, {Youdin}, \& {Simon}}]{LiEtal2019}
{Li}, R., {Youdin}, A.~N., \& {Simon}, J.~B. 2019, \apj, 885, 69

\bibitem[{{Lichtenegger} \& {Komle}(1991)}]{LichteneggerKomle1991}
{Lichtenegger}, H.~I.~M. \& {Komle}, N.~I. 1991, \icarus, 90, 319

\bibitem[{{Lim} {et~al.}(2024){Lim}, {Simon}, {Li}, {Carrera}, {Baronett}, {Youdin}, {Lyra}, \& {Yang}}]{LimEtal2024}
{Lim}, J., {Simon}, J.~B., {Li}, R., {et~al.} 2024, arXiv e-prints, arXiv:2410.17319

\bibitem[{{Lin} {et~al.}(2023){Lin}, {Li}, {Tobin}, {Ohashi}, {J{\o}rgensen}, {Looney}, {Aso}, {Takakuwa}, {Aikawa}, {van't Hoff}, {de Gregorio-Monsalvo}, {Encalada}, {Flores}, {Gavino}, {Han}, {Kido}, {Koch}, {Kwon}, {Lai}, {Lee}, {Lee}, {Phuong}, {Sai}, {Sharma}, {Sheehan}, {Thieme}, {Williams}, {Yamato}, \& {Yen}}]{LinEtal2023}
{Lin}, Z.-Y.~D., {Li}, Z.-Y., {Tobin}, J.~J., {et~al.} 2023, \apj, 951, 9

\bibitem[{{Liu} \& {Ji}(2020)}]{LiuJi2020}
{Liu}, B. \& {Ji}, J. 2020, Research in Astronomy and Astrophysics, 20, 164

\bibitem[{{Liu} {et~al.}(2020){Liu}, {Lambrechts}, {Johansen}, {Pascucci}, \& {Henning}}]{LiuEtal2020}
{Liu}, B., {Lambrechts}, M., {Johansen}, A., {Pascucci}, I., \& {Henning}, T. 2020, \aap, 638, A88

\bibitem[{{Liu} \& {Ormel}(2018)}]{LiuOrmel2018}
{Liu}, B. \& {Ormel}, C.~W. 2018, \aap, 615, A138

\bibitem[{{Liu} {et~al.}(2019){Liu}, {Ormel}, \& {Johansen}}]{LiuEtal2019}
{Liu}, B., {Ormel}, C.~W., \& {Johansen}, A. 2019, \aap, 624, A114

\bibitem[{{Lynden-Bell} \& {Pringle}(1974)}]{Lynden-BellPringle1974}
{Lynden-Bell}, D. \& {Pringle}, J.~E. 1974, \mnras, 168, 603

\bibitem[{{Mathis} {et~al.}(1977){Mathis}, {Rumpl}, \& {Nordsieck}}]{MathisEtal1977}
{Mathis}, J.~S., {Rumpl}, W., \& {Nordsieck}, K.~H. 1977, \apj, 217, 425

\bibitem[{{Morbidelli}(2020)}]{Morbidelli2020i}
{Morbidelli}, A. 2020, \aap, 638, A1

\bibitem[{{Mori} {et~al.}(2019){Mori}, {Bai}, \& {Okuzumi}}]{MoriEtal2019}
{Mori}, S., {Bai}, X.-N., \& {Okuzumi}, S. 2019, \apj, 872, 98

\bibitem[{{Mori} {et~al.}(2021){Mori}, {Okuzumi}, {Kunitomo}, \& {Bai}}]{MoriEtal2021}
{Mori}, S., {Okuzumi}, S., {Kunitomo}, M., \& {Bai}, X.-N. 2021, \apj, 916, 72

\bibitem[{{Musiolik} \& {Wurm}(2019)}]{MusiolikWurm2019}
{Musiolik}, G. \& {Wurm}, G. 2019, \apj, 873, 58

\bibitem[{{Nakagawa} {et~al.}(1986){Nakagawa}, {Sekiya}, \& {Hayashi}}]{NakagawaEtal1986}
{Nakagawa}, Y., {Sekiya}, M., \& {Hayashi}, C. 1986, \icarus, 67, 375

\bibitem[{{{\"O}berg} {et~al.}(2023){{\"O}berg}, {Facchini}, \& {Anderson}}]{OebergEtal2023}
{{\"O}berg}, K.~I., {Facchini}, S., \& {Anderson}, D.~E. 2023, \araa, 61, 287

\bibitem[{{Oka} {et~al.}(2011){Oka}, {Nakamoto}, \& {Ida}}]{OkaEtal2011}
{Oka}, A., {Nakamoto}, T., \& {Ida}, S. 2011, \apj, 738, 141

\bibitem[{{Ormel} \& {Klahr}(2010)}]{OrmelKlahr2010}
{Ormel}, C.~W. \& {Klahr}, H.~H. 2010, \aap, 520, A43

\bibitem[{{Ormel} \& {Liu}(2018)}]{OrmelLiu2018}
{Ormel}, C.~W. \& {Liu}, B. 2018, \aap, 615, A178

\bibitem[{{Ormel} {et~al.}(2017){Ormel}, {Liu}, \& {Schoonenberg}}]{OrmelEtal2017}
{Ormel}, C.~W., {Liu}, B., \& {Schoonenberg}, D. 2017, \aap, 604, A1

\bibitem[{{Ormel} {et~al.}(2015){Ormel}, {Shi}, \& {Kuiper}}]{OrmelEtal2015i}
{Ormel}, C.~W., {Shi}, J.-M., \& {Kuiper}, R. 2015, \mnras, 447, 3512

\bibitem[{{Ormel} {et~al.}(2007){Ormel}, {Spaans}, \& {Tielens}}]{OrmelEtal2007}
{Ormel}, C.~W., {Spaans}, M., \& {Tielens}, A.~G.~G.~M. 2007, \aap, 461, 215

\bibitem[{{Owen}(2020)}]{Owen2020}
{Owen}, J.~E. 2020, \mnras, 495, 3160

\bibitem[{{Ros} \& {Johansen}(2013)}]{RosJohansen2013}
{Ros}, K. \& {Johansen}, A. 2013, \aap, 552, A137

\bibitem[{{Ros} \& {Johansen}(2024)}]{RosJohansen2024}
{Ros}, K. \& {Johansen}, A. 2024, \aap, 686, A237

\bibitem[{{Saito} \& {Sirono}(2011)}]{SaitoSirono2011}
{Saito}, E. \& {Sirono}, S.-i. 2011, \apj, 728, 20

\bibitem[{{Sch{\"a}fer} {et~al.}(2017){Sch{\"a}fer}, {Yang}, \& {Johansen}}]{SchaeferEtal2017}
{Sch{\"a}fer}, U., {Yang}, C.-C., \& {Johansen}, A. 2017, \aap, 597, A69

\bibitem[{{Schoonenberg} {et~al.}(2019){Schoonenberg}, {Liu}, {Ormel}, \& {Dorn}}]{SchoonenbergEtal2019}
{Schoonenberg}, D., {Liu}, B., {Ormel}, C.~W., \& {Dorn}, C. 2019, \aap, 627, A149

\bibitem[{{Schoonenberg} {et~al.}(2017){Schoonenberg}, {Okuzumi}, \& {Ormel}}]{SchoonenbergEtal2017}
{Schoonenberg}, D., {Okuzumi}, S., \& {Ormel}, C.~W. 2017, \aap, 605, L2

\bibitem[{{Schoonenberg} \& {Ormel}(2017)}]{SchoonenbergOrmel2017}
{Schoonenberg}, D. \& {Ormel}, C.~W. 2017, \aap, 602, A21

\bibitem[{{Schoonenberg} {et~al.}(2018){Schoonenberg}, {Ormel}, \& {Krijt}}]{SchoonenbergEtal2018}
{Schoonenberg}, D., {Ormel}, C.~W., \& {Krijt}, S. 2018, \aap, 620, A134

\bibitem[{{Shakura} \& {Sunyaev}(1973)}]{ShakuraSunyaev1973}
{Shakura}, N.~I. \& {Sunyaev}, R.~A. 1973, \aap, 24, 337

\bibitem[{{Simon} {et~al.}(2016){Simon}, {Armitage}, {Li}, \& {Youdin}}]{SimonEtal2016}
{Simon}, J.~B., {Armitage}, P.~J., {Li}, R., \& {Youdin}, A.~N. 2016, \apj, 822, 55

\bibitem[{{Spadaccia} {et~al.}(2022){Spadaccia}, {Capelo}, {Pommerol}, {Schuetz}, {Alibert}, {Ros}, \& {Thomas}}]{SpadacciaEtal2022}
{Spadaccia}, S., {Capelo}, H.~L., {Pommerol}, A., {et~al.} 2022, \mnras, 509, 2825

\bibitem[{{Stone} {et~al.}(2020){Stone}, {Tomida}, {White}, \& {Felker}}]{StoneEtal2020}
{Stone}, J.~M., {Tomida}, K., {White}, C.~J., \& {Felker}, K.~G. 2020, \apjs, 249, 4

\bibitem[{{Takeuchi} \& {Lin}(2002)}]{TakeuchiLin2002}
{Takeuchi}, T. \& {Lin}, D.~N.~C. 2002, \apj, 581, 1344

\bibitem[{{Tobin} {et~al.}(2023){Tobin}, {van't Hoff}, {Leemker}, {van Dishoeck}, {Paneque-Carre{\~n}o}, {Furuya}, {Harsono}, {Persson}, {Cleeves}, {Sheehan}, \& {Cieza}}]{TobinEtal2023}
{Tobin}, J.~J., {van't Hoff}, M. L.~R., {Leemker}, M., {et~al.} 2023, \nat, 615, 227

\bibitem[{{van 't Hoff} {et~al.}(2018){van 't Hoff}, {Tobin}, {Trapman}, {Harsono}, {Sheehan}, {Fischer}, {Megeath}, \& {van Dishoeck}}]{van'tHoffEtal2018}
{van 't Hoff}, M. L.~R., {Tobin}, J.~J., {Trapman}, L., {et~al.} 2018, \apjl, 864, L23

\bibitem[{{Villenave} {et~al.}(2023){Villenave}, {Podio}, {Duch{\^e}ne}, {Stapelfeldt}, {Melis}, {Carrasco-Gonzalez}, {Le Gouellec}, {M{\'e}nard}, {de Simone}, {Chandler}, {Garufi}, {Pinte}, {Bianchi}, \& {Codella}}]{VillenaveEtal2023}
{Villenave}, M., {Podio}, L., {Duch{\^e}ne}, G., {et~al.} 2023, \apj, 946, 70

\bibitem[{{Visser} \& {Ormel}(2016)}]{VisserOrmel2016}
{Visser}, R.~G. \& {Ormel}, C.~W. 2016, \aap, 586, A66

\bibitem[{{Wada} {et~al.}(2013){Wada}, {Tanaka}, {Okuzumi}, {Kobayashi}, {Suyama}, {Kimura}, \& {Yamamoto}}]{WadaEtal2013}
{Wada}, K., {Tanaka}, H., {Okuzumi}, S., {et~al.} 2013, \aap, 559, A62

\bibitem[{{Wang} {et~al.}(2023){Wang}, {Ormel}, {Huang}, \& {Kuiper}}]{WangEtal2023}
{Wang}, Y., {Ormel}, C.~W., {Huang}, P., \& {Kuiper}, R. 2023, \mnras, 523, 6186

\bibitem[{{Weidenschilling}(1977)}]{Weidenschilling1977}
{Weidenschilling}, S.~J. 1977, \mnras, 180, 57

\bibitem[{{Xu} \& {Kunz}(2021)}]{XuKunz2021}
{Xu}, W. \& {Kunz}, M.~W. 2021, \mnras, 508, 2142

\bibitem[{{Xu} {et~al.}(2017){Xu}, {Bai}, \& {Murray-Clay}}]{XuEtal2017}
{Xu}, Z., {Bai}, X.-N., \& {Murray-Clay}, R.~A. 2017, \apj, 847, 52

\bibitem[{{Yang} {et~al.}(2018){Yang}, {Mac Low}, \& {Johansen}}]{YangEtal2018}
{Yang}, C.-C., {Mac Low}, M.-M., \& {Johansen}, A. 2018, \apj, 868, 27

\bibitem[{{Youdin} \& {Goodman}(2005)}]{YoudinGoodman2005}
{Youdin}, A.~N. \& {Goodman}, J. 2005, \apj, 620, 459

\bibitem[{{Youdin} \& {Lithwick}(2007)}]{YoudinLithwick2007}
{Youdin}, A.~N. \& {Lithwick}, Y. 2007, \icarus, 192, 588

\bibitem[{{Zhang}(2024)}]{Zhang2024}
{Zhang}, K. 2024, Reviews in Mineralogy and Geochemistry, 90, 27

\bibitem[{{Zhu} {et~al.}(2015){Zhu}, {Stone}, \& {Bai}}]{ZhuEtal2015}
{Zhu}, Z., {Stone}, J.~M., \& {Bai}, X.-N. 2015, \apj, 801, 81

\end{thebibliography}

\begin{appendix} 
	\section{Lagrangian particle tracking}
	\label{sec:lagrangian}
	Our approach to implement the Lagrangian particle tracking follows \citet{Ciesla2010}.
	After a timestep $\Delta t$ the positions of water particles are updated through,
	\begin{equation}
		(r,z)_{t+\Delta t} = (r,z)_{t} + \bm{v}_{\mathrm{eff}} \Delta t + R \left[\frac{2}{\zeta} D \Delta t \right]^{1/2}
	\end{equation}
	with $\zeta = 1/3$, $D$ the diffusivity of water in either ice or vapor state (we drop the subscript ``p'' or ``g'' hereafter for simplicity) and $R$ a random number between $[-1,1]$ following uniform distribution. The effective velocity $\bm{v}_{\mathrm{eff}} = \left(v_{\mathrm{eff},r},v_{\mathrm{eff},z}\right)$ has only its $z$-component given in \cite{Ciesla2010}. Here we re-derive it to obtain also the $r$-component.

	The dynamical evolution of water parcels follows the diffusion-advection equation,
	\begin{equation}
		\frac{\partial \rho}{\partial t} + \nabla \cdot (\rho \mathbf{v} - D \rho_{\gas} \nabla c ) = 0,
	\end{equation}
	where $c = \rho / \rho_{\gas}$ is the concentration of either ice or vapor.
	In cylindrical coordinate, the equation reads,
	\begin{equation}
		\label{eq:DA_cyl}
		\frac{\partial \rho}{\partial t} + \frac{1}{r} \frac{\partial}{\partial r} (r v_{r} \rho) + \frac{\partial}{\partial z} (v_{z} \rho) = \frac{1}{r} \frac{\partial}{\partial r} \left( \rho_{\gas} D r \frac{\partial c}{\partial r} \right) + \frac{\partial}{\partial z} \left(\rho_{\gas} D \frac{\partial c}{\partial z}\right).
	\end{equation}
	To reproduce the dynamical evolution of water parcels subject to diffusive-advective motions in the way of Monte Carlo tracking of particles, the Monte Carlo methods should satisfy the Fokker-Planck equation:
	\begin{equation}
		\frac{\partial f}{\partial t} + \sum_{i=1}^{N} \frac{\partial}{\partial x_{i}} \left( v_{i} f \right) = \sum_{i=1}^{N} \sum_{j=1}^{N} \frac{\partial^{2}}{\partial x_{i} \partial x_{j}} \left( D_{ij} f \right),
	\end{equation}
	where $i,j$ denote different dimensions and $N=2$ in our case. The Fokker-Planck equation governs the evolution of the probability distribution of the variable $\rho$, which is denoted as $f$ in above equation. In the simulation we assume $D_{ij} = D(r)$, then the equation becomes:
	\begin{equation}
		\label{eq:FP}
		\begin{aligned}
			\frac{\partial f}{\partial t} & + \frac{1}{r} \frac{\partial}{\partial r} (r v_{r} f) + \frac{\partial}{\partial z} (v_{z} f) = \frac{1}{r} \frac{\partial}{\partial r} \left[r \frac{\partial (D f)}{\partial r} \right] + \frac{\partial^{2}}{\partial z^{2}} \left(D f\right) \\
			                              & \equiv \frac{1}{r} \frac{\partial}{\partial r} \left[r \frac{\partial D}{\partial r} f + r D \frac{\partial f}{\partial r} \right] + \frac{\partial^{2}}{\partial z^{2}} \left(D f\right).
		\end{aligned}
	\end{equation}
	To rewrite the diffusion-advection equation into the form of Fokker-Planck equation, we expand \eq{DA_cyl} as,
	\begin{equation}
		\label{eq:AD_expand}
		\begin{aligned}
			\frac{\partial \rho}{\partial t} & + \frac{1}{r} \frac{\partial}{\partial r} (r v_{r} \rho) + \frac{\partial}{\partial z} (v_{z} \rho) = \frac{1}{r} \frac{\partial}{\partial r} \left(rD \frac{\partial \rho}{\partial r} \right)                                                                                                      \\
			                                 & - \frac{1}{r} \frac{\partial}{\partial r} \left(r\frac{D}{\rho_{\gas}} \frac{\partial\rho_{\gas}}{\partial r} \rho \right) + \frac{\partial^{2}}{\partial z^{2}} \left(D \rho \right) - \frac{\partial}{\partial z} \left(\frac{D}{\rho_{\gas}} \frac{\partial\rho_{\gas}}{\partial z} \rho \right).
		\end{aligned}
	\end{equation}
	Comparing \eqs{FP}{AD_expand}, we can get,
	\begin{equation}
		\begin{aligned}
			v_{\mathrm{eff},r} & = v_{r} + \frac{D}{\rho_{\gas}} \frac{\partial \rho_{\gas}}{\partial r} + \frac{\partial D}{\partial r}, \\
			v_{\mathrm{eff},z} & = v_{z} +  \frac{D}{\rho_{\gas}} \frac{\partial \rho_{\gas}}{\partial z}.
		\end{aligned}
	\end{equation}
	The second terms in both $v_{\mathrm{eff},r}$ and $v_{\mathrm{eff},z}$ comes from density stratification, that particles are more likely to appear near denser region. The last term in $v_{\mathrm{eff},r}$ comes from the radial gradient of diffusivity. In the Monte Carlo simulation, $v_{r}$ and $v_{z}$ are taken from steady state velocity field of either ice and vapor in the hydrodynamic simulations.

	Next, the phase change probability is determined according to the sublimation rate ($R_{\mathrm{e}}$) and condensation rate ($R_{\mathrm{c}}$). Following \citet{RosJohansen2013}, we can define $R_{\mathrm{e}}$ and $R_{\mathrm{c}}$ as
	\begin{equation}
		\frac{dm}{dt} = 4 \pi R^{2} v_{\mathrm{th}} \rho_{\mathrm{vap}} \left( 1-\frac{P_{\mathrm{eq}}}{P_{\mathrm{vap}}}  \right) \equiv R_{\mathrm{c}} \rho_{\mathrm{vap}} - R_{\mathrm{e}} \rho_{\mathrm{p,ice}},
	\end{equation}
	where $m$ is the mass of a pebble. Therefore,
	\begin{equation}
		\frac{R_{\mathrm{c}}}{R_{\mathrm{e}}} = \frac{P_{\mathrm{vap}}}{P_{\mathrm{eq}}} \frac{\rho_{\mathrm{p,ice}}}{\rho_{\mathrm{vap}}}.
	\end{equation}
	And the probability to have phase change happening for Lagrangian particles follows as,
	\begin{equation}
		P = \left\{
		\begin{aligned}
			R_{\mathrm{e}}/(R_{\mathrm{e}} + R_{\mathrm{c}}) & , & (\mathrm{sublimation}),  \\
			R_{\mathrm{c}}/(R_{\mathrm{e}} + R_{\mathrm{c}}) & , & (\mathrm{condensation}).
		\end{aligned}
		\right.
	\end{equation}

	In \fg{trace_contour} we plot $\rho_{\mathrm{p,ice}} / \rho_{\mathrm{vap}}$ to denote the respective probability to be either in ice or vapor phase since $R_{\mathrm{c}} / R_{\mathrm{e}}$ closely follows $\rho_{\mathrm{p,ice}} / \rho_{\mathrm{vap}}$ at the region where phase change happens frequently.

	\section{Update of 1D snowline model in gas dynamics}
	\label{sec:1D_model}
	To verify our hydrodynamic model, we simulate the 1D snowline following the set-up in \citet{SchoonenbergOrmel2017}. The disk temperature profile is fixed as,
	\begin{equation}
		T(r) = 150 \left(\frac{r}{3~\mathrm{au}}\right)^{-1/2}~\mathrm{K}.
	\end{equation}
	The gas is supplied at the outer boundary at a constant rate $\dot{M}_{\mathrm{acc}} = 10^{-8} M_{\odot} ~\mathrm{Yr}^{-1}$. Other parameters ($f_{\mathrm{i/g}}$, $\tau_{\mathrm{s,0}}$, $\alpha$) are also kept the same as 2D runs (\tb{simu_cases}). In this way, the disk snowline is expected to be located at $\sim 2.1~\mathrm{au}$. In previous studies \citep{SchoonenbergOrmel2017,HyodoEtal2021}, the gas and dust motion are solved with a set of transport equations, where the gas velocity and the surface density of non-condensible gas ($\Sigma_{\mathrm{xy}}$) are fixed as the viscous solution of an unperturbed disk composed of pure H-He,
	\begin{equation}
		\label{eq:simple_model}
		v_{\mathrm{g}} = \frac{3 \nu}{2 r}; \quad \Sigma_{\mathrm{xy}} = \frac{\dot{M}_{\mathrm{acc}}}{3 \pi \nu}.
	\end{equation}
	This treatment ignores two important aspects which would reshape the solid pile-up.
	Firstly, with ice sublimation, the surface density of the gas is altered with injected vapor and the gas velocity should deviate accordingly following (e.g., \citealt{Lynden-BellPringle1974}),
	\begin{equation}
		v_{\mathrm{g}} = -\frac{3}{r \Sigma v_{\mathrm{k}}} \frac{\partial}{\partial r} \left( r \Sigma v_{\mathrm{k}} \nu \right).
	\end{equation}
	Secondly, as vapor diffusion occurs, the H-He gas also diffuses in a similar manner. Consequently, $\Sigma_{\mathrm{xy}}$ no longer sticks to the pure advection solution. Instead, it must be determined by the combined effects of advection and diffusion, as described by the advection-diffusion process (see \eq{CE_vapor}).

	By incorporating the momentum equation and solving for the diffusion of all species within the hydrodynamic model, the changes in gas velocity and the effects of H-He diffusion are inherently accounted for. To demonstrate their effect on the pile-up, a ``simple model'' (\texttt{1D/SO17}) is constructed, where both the gas velocity and $\Sigma_{\mathrm{xy}}$ are fixed following \eq{simple_model}. In \fg{1D_model}, we compare the pile-up strength of the ``simple'' model and the contrasting ``full'' model (\texttt{1D-hydro}). As expected, the surface density of H-He gas ($\Sigma_{\mathrm{xy}}$) deviates from the unperturbed viscous solution, manifesting a density bump outside the snowline (\fg{1D_model}a). Importantly, the \texttt{1D-hydro} run presents around a factor of two's increase in the peak solid-to-gas ratio (\fg{1D_model}b). These can be understood as the result of midplane viscous outflow of gas near the snowline, which significantly slows down the drift of pebbles (\fg{1D_model}c) and leads to a stronger pile-up. The 1D results therefore confirm the importance of gas dynamics in shaping the solid pile-up outside the snowline, which is already discussed in \se{gas_dynamics}.

	\begin{figure}
		\includegraphics[width=\columnwidth]{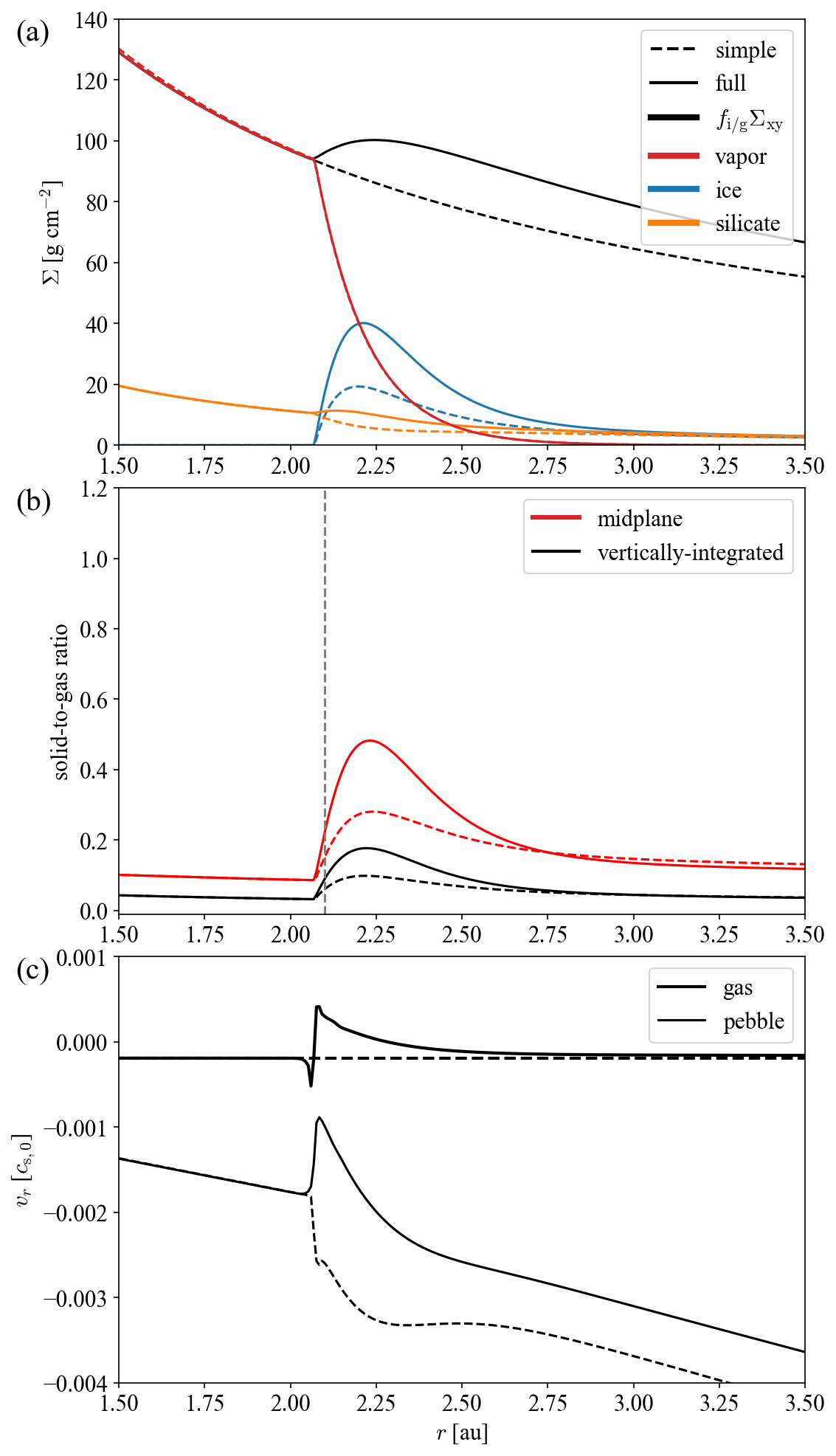}
		\caption{Comparisons of \texttt{1D/SO17} and \texttt{1D-hydro}. \textbf{(a)}: The surface density profiles. The H-He surface density is timed by the ice-to-gas ratio $f_{\mathrm{i/g}}$ for a clear comparison. \textbf{(b)}: The solid-to-gas ratio at the disk midplane and the vertically-integrated surface density ratio. The grey dashed line denotes snowline location (2.1 au). \textbf{(c)}: The gas and pebble radial velocity at the disk midplane. It is worth noticing that the gas velocity in \texttt{1D/SO17} haven't returned to the unperturbed viscous solution even at the outer boundary, giving rise to the difference of $\Sigma_{\mathrm{xy}}$ all over the place outside the snowline. }
		\label{fig:1D_model}
	\end{figure}

	\section{Effect of disk viscosity}
	\label{sec:vary_alpha}
	\md{To illustrate the effect of viscosity, we vary the disk viscosity parameter $\alpha$ to $0.03$ and $0.001$ with respect to the fiducial value $0.003$ (we do not choose $\alpha=3\times 10^{-4}$ since it takes too much computational effort to reach a steady state). Here we only study the active disk case, where the water cycle is vigorous. Following the design in \se{simu_cases}, the initial snowline is anchored at 2.1 au, therefore $\kappa_{\mathrm{R}}$ is adjusted accordingly. All other parameters (e.g., $\dot{M}_{\mathrm{acc}}$ and $f_{\mathrm{i/g}}$) are kept the same as in the \texttt{active} run (see \se{methods} and \tb{simu_cases}). The steady state radial profiles are presented in \fg{profile_vary_alpha} and the density contours in \fg{contourf_vary_alpha}.
	}

	\md{First, the water cycle persists with various disk viscosity, as can be seen from the settling of pebbles in \fg{contourf_vary_alpha}.}
	\md{Second, as $\alpha$ increases, the water cycle extends over a larger region in the $r-z$ plane. This extension is also clearly seen in surface density profiles of ice, where the pile-ups become wider with increasing $\alpha$ (\fg{profile_vary_alpha}). This occurs because (1) the increased scale height of pebbles allows vapor to recondense at higher altitudes, and (2) the stronger gas background outflow in the midplane — driven by higher disk viscosity (e.g., \citealt{TakeuchiLin2002}) — impedes pebble drift. Comparing the movement of pebbles in \fg{contourf_vary_alpha}a,b and \fg{contourf_density}b, we observe that despite a fixed Stokes number, pebbles drift more efficiently at lower $\alpha$, whereas at $\alpha = 0.03$, they are strongly influenced by gas motion and can even move outward near the midplane.} \md{As a result, a higher $\alpha$ leads to much higher level of solid pile-up in terms of both the amount of ice accumulated and the solid-to-gas ratio (\fg{profile_vary_alpha}).}

	\md{In the case of $\alpha=0.03$, the strong outflow not only promotes the pile-up of ice, but also leads to the pile-up of silicate. As shown in \fg{profile_vary_alpha} (left panel), the surface density of silicate at the snowline becomes comparable to that of ice. This arises due to a traffic jam effect induced by the gas outflow, which slows silicate drift near the snowline. Consequently, predictions based purely on the vapor retro-diffusion scenario may significantly underestimate the strength of solid pile-up. For instance, in steady state, our simulation yields $\xi_{\mathrm{pk}} \approx 1.5$ for $\alpha=0.03$, whereas \citet{SchoonenbergOrmel2017} reports $\xi_{\mathrm{pk}} \approx 0.8$ (see their Fig. 8). Future works should take into account the effect of gas hydrodynamics, especially at high viscosity.}

	\begin{figure*}
		\centering
		\includegraphics[width=1.95\columnwidth]{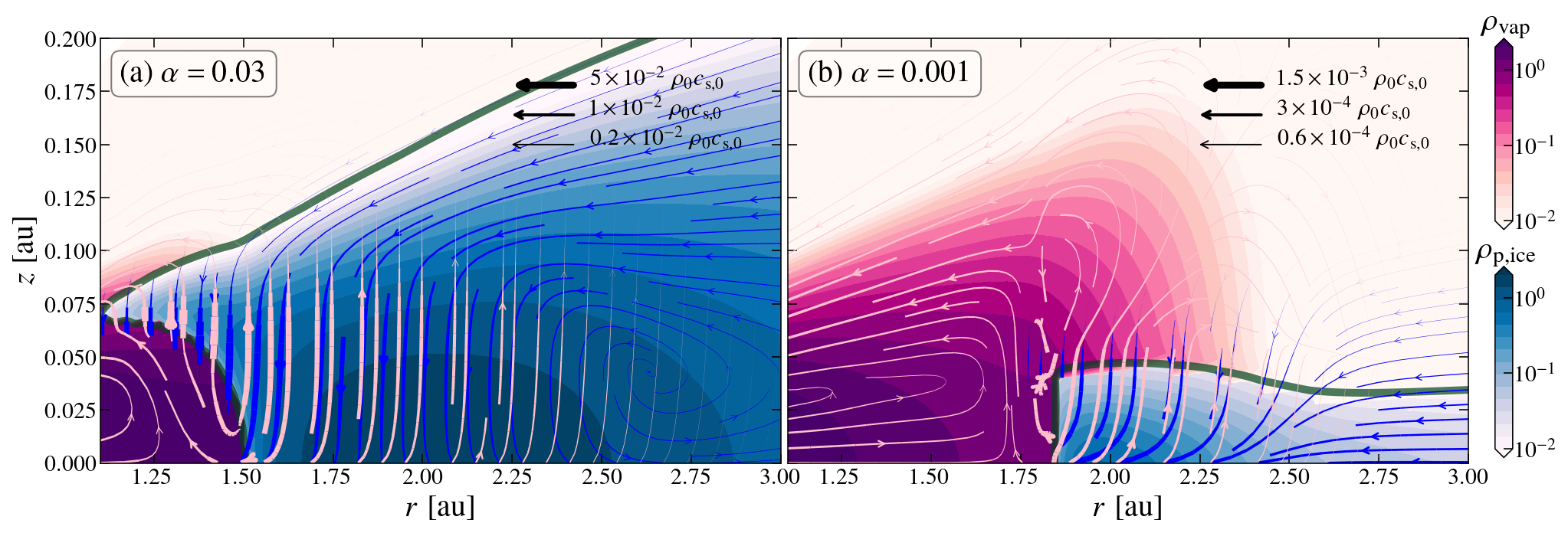}
		\caption{Steady state density structure of active disk runs of different $\alpha$. The notation is the same as \fg{contourf_density} except for different $\alpha$, $\rho_{0}$ (the midplane gas density at $r_{0}$ of the unperturbed disk) is different ($\rho_{0} \approx 3.4 \times 10^{-12} ~\mathrm{cm~g^{-3}}$ for $\alpha = 0.03$ and $\approx 1.0 \times 10^{-10} ~\mathrm{cm~g^{-3}}$ for $\alpha = 0.001$).}
		\label{fig:contourf_vary_alpha}
	\end{figure*}

	\begin{figure*}
		\centering
		\includegraphics[width=1.95\columnwidth]{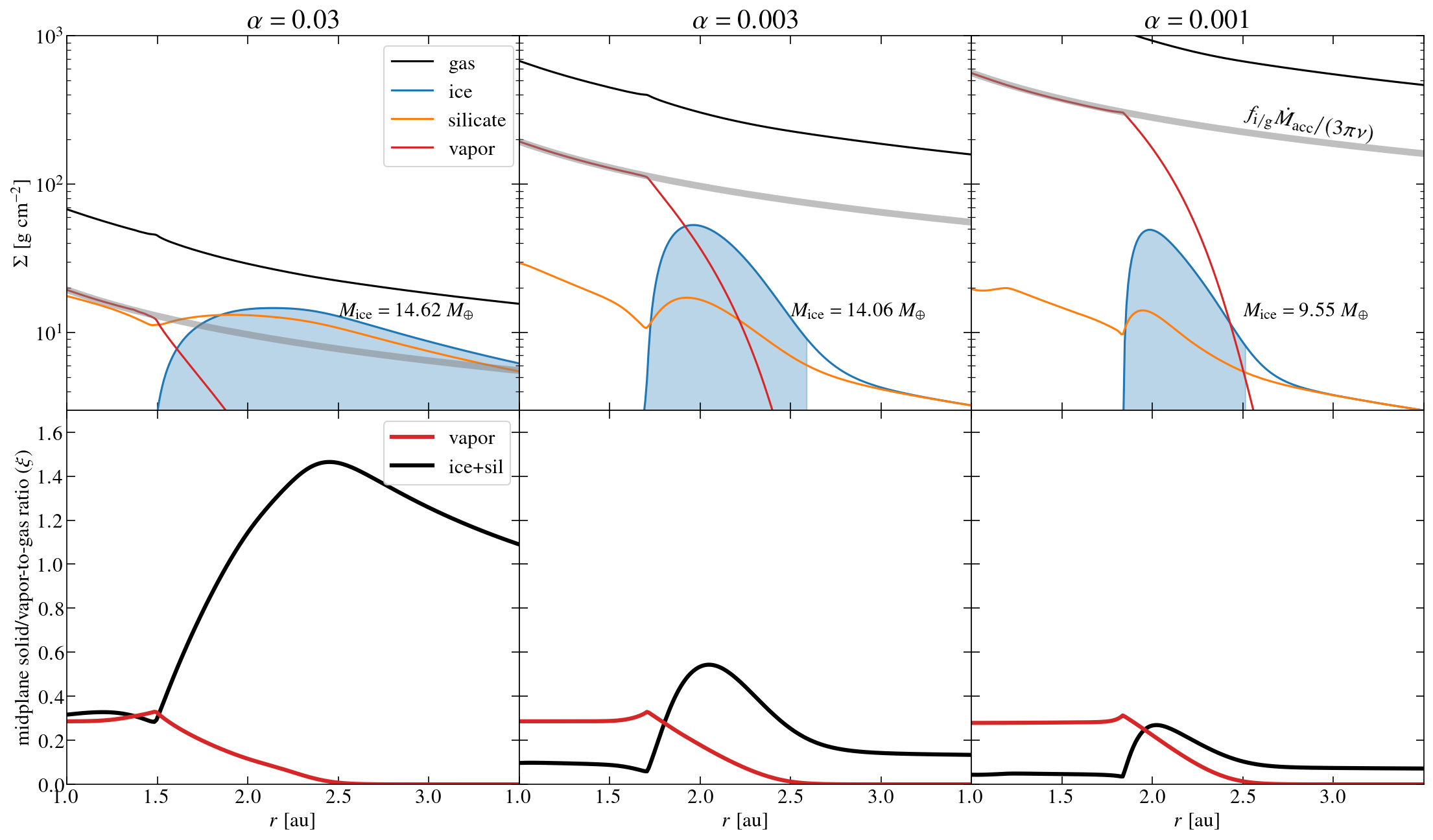}
		\caption{Steady state radial profiles of active disk runs of different $\alpha$. The notation is the same as \fg{profile_iso_highL}. Given the fixed $\dot{M}_{\mathrm{acc}}$, the gas and surface density is much larger when $\alpha = 0.001$ since they are accreted slower. }
		\label{fig:profile_vary_alpha}
	\end{figure*}
\end{appendix}

\end{document}